\documentclass{article}
\usepackage{graphicx,amsmath,amssymb,amsfonts,hyperref}
\begin{document}

\title{Crowd Behavior Dynamics: Entropic Path--Integral Model}
\author{Vladimir G. Ivancevic, Darryn J. Reid, and Eugene V. Aidman\\
\small{Land Operations Division, Defence Science \& Technology Organisation}\\ \small{Adelaide, Australia}}\date{}\maketitle

\begin{abstract}
We propose an entropic geometrical model of crowd behavior dynamics (with dissipative crowd kinematics), using Feynman action--amplitude formalism that operates on three synergetic levels: macro, meso and micro. The intent is to explain the dynamics of crowds simultaneously and consistently across these three levels, in order to characterize their geometrical properties particularly with respect to behavior regimes and the state changes between them. Its most natural statistical descriptor (order parameter) is crowd entropy $S$ that satisfies the Prigogine's extended second law of thermodynamics, $\partial_tS\geq 0$ (for any nonisolated multi-component system). Qualitative similarities and superpositions between individual and crowd configuration manifolds motivate our claim that \emph{goal-directed crowd movement} operates under entropy conservation, $\partial_tS = 0$, while \emph{naturally chaotic crowd dynamics} operates under (monotonically) increasing entropy function, $\partial_tS > 0$. Between these two distinct topological phases lies a phase transition with a chaotic inter-phase. Both inertial crowd dynamics and its dissipative kinematics represent diffusion processes on the crowd manifold governed by the Ricci flow.\\
\\
\textbf{Keywords:}
Crowd behavior dynamics, action--amplitude formalism, entropic crowd manifold, crowd turbulence, Ricci flow, topological phase transitions.
\end{abstract}


\section{Introduction}

Today it is well known that disembodied cognition is a myth, albeit one that has had profound influence in Western science since Rene Descartes and others gave it credence during the Scientific Revolution. In fact, the mind-body separation had much more to do with explanation of method than with explanation of the mind and cognition, yet it is with respect to the latter that its impact is most widely felt. We find it to be an unsustainable assumption in the realm of crowd behavior. Mental intention is
(almost immediately) followed by a physical action, that is, a human or
animal movement \cite{Schoner}. In animals, this physical action would be
jumping, running, flying, swimming, biting or grabbing. In humans, it can be
talking, walking, driving, or kicking, etc. Mathematical description of
human/animal movement in terms of the corresponding neuro-musculo-skeletal
equations of motion, for the purpose of prediction and control, is
formulated within the realm of biodynamics (see \cite%
{GaneshIEEE,IJMMS1,SIAM,VladNick,LieLagr,GaneshSprSml,GaneshWSc,GaneshSprBig}).

The crowd (or, collective) behavior dynamics is clearly formed by
some kind of \textit{superposition, contagion, emergence,} or \textit{convergence} from the individual agents' behavior. According to the emergence theory
\cite{Turner}, crowds begin as collectivities composed of people with mixed
interests and motives; especially in the case of less stable crowds
(expressive, acting and protest crowds) norms may be vague and changing;
people in crowds make their own rules as they go along. According to currently popular convergence theory (see \cite{HelbingPRE1,Nara}), crowd behavior is not a product of the crowd itself, but is carried into the crowd by particular individuals, thus crowds amount to a convergence of like–minded individuals.

We propose that the contagion and convergence theories may be unified by acknowledging that both factors may coexist, even within a single scenario: we propose to refer to this third approach as \emph{behavioral composition}. It represents a substantial shift from traditional analytical approaches, which have assumed either reduction of a whole into parts or the emergence of the whole from the parts. In particular, both contagion and convergence are related to social entropy, which is the natural decay of structure (such as law, organization, and convention) in a social system \cite{Downarowicz}.

In this paper we attempt to formulate a geometrically predictive model--theory of crowd behavior dynamics, based on the previously formulated individual Life Space Foam concept \cite{IA} (see Appendix for the brief summary).

It is today well known that massive crowd movements can be precisely observed/monitored from satellites and all that one can see is crowd physics. Therefore, all involved psychology of individual crowd agents: cognitive, motivational and emotional -- is only a non-transparent input (a hidden initial switch) for the fully observable crowd physics. In this paper we will label this initial switch as `mental preparation' or `loading', while the manifested physical action is labeled `execution'. We propose the entropy formulation of crowd dynamics as a three--step process involving individual behavior dynamics and collective behavior dynamics. The chaotic behavior phase-transitions embedded in crowd dynamics may give a formal description for a phenomenon called \emph{crowd turbulence} by D. Helbing, depicting crowd disasters caused by the panic stampede that can occur at high pedestrian densities and which is a serious concern during mass events like soccer championship games or annual pilgrimage in Makkah (see \cite{HelbingNature,HelbingPRL,HelbingPRE,HelbingACS}).

\section{Generic three--step crowd behavior dynamics}

In this section we propose a generic crowd behavior dynamics as a three--step process based on a general partition function formalism. Note that the number of variables $X_{i}$ in the standard \emph{partition function} from statistical mechanics (see, e.g. \cite{Landau}) need not be countable, in which case
the set of coordinates $\{x^{i}\}$ becomes a field\ $\phi =\phi (x)$. The sum is replaced by the \emph{Euclidean path integral} (that is a
Wick--rotated Feynman transition amplitude in imaginary time, see subsection \ref{aggreg}), as
\begin{equation*}
Z(\phi )=\int \mathcal{D}[\phi ]\exp \left[ -H(\phi )\right] .
\end{equation*}
More generally, in quantum field theory, instead of the field Hamiltonian $%
H(\phi )$ we have the action $S(\phi )$ of the theory. Both Euclidean path
integral,
\begin{equation}
Z(\phi )=\int \mathcal{D}[\phi ]\exp \left[ -S(\phi )\right] ,\qquad \text{%
real path integral in imaginary time}  \label{Eucl}
\end{equation}%
and Lorentzian one,
\begin{equation}
Z(\phi )=\int \mathcal{D}[\phi ]\exp \left[ iS(\phi )\right] ,\qquad \text{%
complex path integral in real time}  \label{Lor}
\end{equation}%
-- represent quantum field theory (QFT) partition functions. We will give formal definitions of the above path integrals (i.e., general partition functions) in section 3. For the moment, we only remark that the Lorentzian path integral
(\ref{Lor}) gives a QFT generalization of the (nonlinear) Schr\"{o}dinger equation, while the Euclidean path integral (\ref{Eucl}) in the (rectified) real time represents a
statistical field theory (SFT) generalization of the Fokker--Planck equation.

Now, following the framework of the Prigogine's Extended Second Law of Thermodynamics \cite{Nicolis}, $\partial _{t}S\geq 0,$ for entropy $S$ in any complex
system described by its partition function, we formulate a generic crowd behavior dynamics, based on above partition functions, as the
following three--step process:

1. Individual behavior dynamics ($\mathcal{ID}$) is a transition
process from an entropy--growing \textquotedblleft loading" phase of mental
preparation, to the entropy--conserving \textquotedblleft execution"
phase of physical action. Formally, $\mathcal{ID}$ is given by the
phase-transition map:
\begin{equation}
\mathcal{ID}:~\overset{``\mathrm{LOADING}":\,\partial _{t}S>0}{\overbrace{%
\mathrm{MENTAL~PREPARATION}}}~\Longrightarrow ~\overset{``\mathrm{EXECUTION}%
":\,\partial _{t}S=0}{\overbrace{\mathrm{PHYSICAL~ACTION}}}  \label{id}
\end{equation}%
defined by the individual (chaotic) phase-transition amplitude
\begin{equation*}
\left\langle \overset{\partial _{t}S=0}{\mathrm{PHYS.~ACTION}}\right\vert
CHAOS\left\vert \overset{\partial _{t}S>0}{\mathrm{MENTAL~PREP.}}%
\right\rangle _{\mathrm{ID}}:=\int \mathcal{D}[\Phi ]\,\mathrm{e}^{iS_{%
\mathrm{ID}}[\Phi ]},
\end{equation*}%
where the right-hand-side is the Lorentzian path-integral (or complex
path-integral in real time), with the individual behavior action
\begin{equation*}
S_{\mathrm{ID}}[\Phi ]=\int_{t_{ini}}^{t_{fin}}{L}_{\mathrm{ID}}[\Phi ]\,dt,
\end{equation*}%
where ${L}_{\mathrm{ID}}[\Phi ]$ is the \emph{behavior Lagrangian,}
consisting of \emph{mental cognitive potential} and physical \emph{kinetic energy}.

2. Aggregate behavior dynamics ($\mathcal{AD}$) represents the
behavioral composition--transition map:
\begin{equation}
\mathcal{AD}:~\overset{``\mathrm{LOADING}":\,\partial _{t}S>0}{\sum_{i\in
\mathrm{AD}}\overbrace{\mathrm{MENTAL~PREPARATION}}}~\Longrightarrow
\sum_{i\in \mathrm{AD}}\overset{``\mathrm{EXECUTION}":\,\partial _{t}S=0}{%
\overbrace{\mathrm{PHYSICAL~ACTION}}}_{i}  \label{ad}
\end{equation}%
where the (weighted) aggregate sum is taken over all individual agents,
assuming equipartition of the total behavioral energy. It is defined
by the aggregate (chaotic) phase-transition amplitude
\begin{equation*}
\left\langle \overset{\partial _{t}S=0}{\mathrm{PHYS.~ACTION}}\right\vert
CHAOS\left\vert \overset{\partial _{t}S>0}{\mathrm{MENTAL~PREP.}}%
\right\rangle _{\mathrm{AD}}:=\int \mathcal{D}[\Phi ]\,\mathrm{e}^{-S_{%
\mathrm{AD}}[\Phi ]},
\end{equation*}%
with the Euclidean path-integral in real time,
that is the SFT--partition function, based on the
aggregate behavior action
\begin{equation*}
S_{\mathrm{AD}}[\Phi ]=\int_{t_{ini}}^{t_{fin}}{L}_{\mathrm{AD}}[\Phi
]\,dt,\qquad \mathrm{with}\qquad {L}_{\mathrm{AD}}[\Phi ]=\sum_{i\in \mathrm{%
AD}}{L}_{\mathrm{ID}}^{i}[\Phi ].
\end{equation*}

3. Crowd behavior dynamics ($\mathcal{CD}$) represents the
cumulative transition map:
\begin{equation}
\mathcal{CD}:~\overset{``\mathrm{LOADING}":\,\partial _{t}S>0}{\sum_{i\in
\mathrm{CD}}\overbrace{\mathrm{MENTAL~PREPARATION}}}~\Longrightarrow
\sum_{i\in \mathrm{CD}}\overset{``\mathrm{EXECUTION}":\,\partial _{t}S=0}{%
\overbrace{\mathrm{PHYSICAL~ACTION}}}_{i}  \label{cd}
\end{equation}%
where the (weighted) cumulative sum is taken over all individual agents,
assuming equipartition of the total behavior energy. It is defined
by the crowd (chaotic) phase-transition amplitude
\begin{equation*}
\left\langle \overset{\partial _{t}S=0}{\mathrm{PHYS.~ACTION}}\right\vert
CHAOS\left\vert \overset{\partial _{t}S>0}{\mathrm{MENTAL~PREP.}}%
\right\rangle _{\mathrm{CD}}:=\int \mathcal{D}[\Phi ]\,\mathrm{e}^{iS_{%
\mathrm{CD}}[\Phi ]},
\end{equation*}%
with the general Lorentzian path-integral, that is, the QFT--partition function), based on the crowd behavior action
\begin{equation*}
S_{\mathrm{CD}}[\Phi ]=\int_{t_{ini}}^{t_{fin}}{L}_{\mathrm{CD}}[\Phi
]\,dt,\qquad \mathrm{with}\qquad {L}_{\mathrm{CD}}[\Phi ]=\sum_{i\in \mathrm{%
CD}}{L}_{\mathrm{ID}}^{i}[\Phi ]=\sum_{k=\,\text{\# of }\mathrm{ADs}%
\text{ in CD}}{L}_{\mathrm{AD}}^{k}[\Phi ].
\end{equation*}

All three entropic phase-transition maps, $\mathcal{ID}$, $\mathcal{AD}$ and $\mathcal{CD}$, are
spatio--temporal biodynamic cognition systems \cite{CompMind}, evolving within their
respective configuration manifolds (i.e., sets of their respective degrees-of-freedom with equipartition of energy), according to biphasic
action--functional formalisms with behavior--Lagrangian functions $L_{%
\mathrm{ID}}$, $L_{\mathrm{AD}}$ and $L_{\mathrm{CD}}$, each consisting of:
\begin{enumerate}
\item Cognitive mental potential (which is a mental preparation for the
physical action), and

\item Physical kinetic energy (which describes the physical action itself).
\end{enumerate}

To develop $\mathcal{ID}$, $\mathcal{AD}$ and $\mathcal{CD}$ formalisms, we extend into a
physical (or, more precisely, biodynamic) crowd domain a purely--mental individual
Life--Space Foam (LSF) framework for motivational cognition \cite{IA}, based
on the quantum--probability concept.\footnote{The quantum probability concept is based on the following physical facts
\cite{Complexity,QuLeap}
\begin{enumerate}
\item {The time--dependent Schr\"{o}dinger equation} represents a {%
complex--valued generalization} of the real--valued {Fokker--Planck equation}
for describing the spatio--temporal {probability density function} for the
system exhibiting {continuous--time Markov stochastic process}.

\item The Feynman path integral (including integration over continuous
spectrum and summation over discrete spectrum) is a generalization of the
time--dependent Schr\"{o}dinger equation, including both continuous--time
and discrete--time Markov stochastic processes.

\item Both Schr\"{o}dinger equation and path integral give `physical
description' of any system they are modelling in terms of its physical
energy, instead of an abstract probabilistic description of the
Fokker--Planck equation.
\end{enumerate}
Therefore, the Feynman path integral, as a generalization of the (nonlinear)
time--dependent Schr\"{o}dinger equation, gives a unique physical
description for the general Markov stochastic process, in terms of the
physically based generalized probability density functions, valid both for
continuous--time and discrete--time Markov systems. Its basic consequence is this: a different way for calculating probabilities. The
difference is rooted in the fact that \textsl{sum of squares is different
from the square of sums}, as is explained in the following text. Namely, in
Dirac--Feynman quantum formalism, each possible route from the initial
system state $A$ to the final system state $B$ is called a {history}. This
history comprises any kind of a route, ranging from continuous and smooth
deterministic (mechanical--like) paths to completely discontinues and random
Markov chains (see, e.g., \cite{Gardiner}). Each history (labelled by index $%
i$) is quantitatively described by a {complex number}.

In this way, the overall probability of the system's transition from some
initial state $A$ to some final state $B$ is given {not} by adding up the
probabilities for each history--route, but by `head--to--tail' adding up the
sequence of amplitudes making--up each route first (i.e., performing the
sum--over--histories) -- to get the total amplitude as a `resultant vector',
and then squaring the total amplitude to get the overall transition
probability.

Here we emphasize that the domain of validity of the `quantum' is not
restricted to the microscopic world \cite{Ume93}. There are macroscopic
features of classically behaving systems, which cannot be explained without
recourse to the quantum dynamics. This field theoretic model leads to the
view of the phase transition as a condensation that is comparable to the
formation of fog and rain drops from water vapor, and that might serve to
model both the gamma and beta phase transitions. According to such a model,
the production of activity with long--range correlation in the brain takes
place through the mechanism of spontaneous breakdown of symmetry (SBS),
which has for decades been shown to describe long-range correlation in
condensed matter physics. The adoption of such a field theoretic approach
enables modelling of the whole cerebral hemisphere and its hierarchy of
components down to the atomic level as a fully integrated macroscopic
quantum system, namely as a macroscopic system which is a quantum system not
in the trivial sense that it is made, like all existing matter, by quantum
components such as atoms and molecules, but in the sense that some of its
macroscopic properties can best be described with recourse to quantum
dynamics (see \cite{FreVit06} and references therein). Also, according to
Freeman and Vitielo, \textit{many--body quantum field theory} appears to be
the only existing theoretical tool capable to explain the dynamic origin of
long--range correlations, their rapid and efficient formation and
dissolution, their interim stability in ground states, the multiplicity of
coexisting and possibly non--interfering ground states, their degree of
ordering, and their rich textures relating to sensory and motor facets of
behaviors. It is historical fact that many--body quantum field theory has
been devised and constructed in past decades exactly to understand features
like ordered pattern formation and phase transitions in condensed matter
physics that could not be understood in classical physics, similar to those
in the brain.}

The behavioral approach to $\mathcal{ID}$, $\mathcal{AD}$ and $\mathcal{CD}$ is based
on \textit{entropic motor control} \cite{Hong1,Hong2}, which deals with
neuro-physiological feedback information and environmental uncertainty. The
probabilistic nature of human motor action can be characterized by entropies
at the level of the organism, task, and environment. Systematic changes in
motor adaptation are characterized as task--organism and
environment--organism tradeoffs in entropy. Such compensatory adaptations
lead to a view of goal--directed motor control as the product of an
underlying conservation of entropy across the task--organism--environment
system. In particular, an experiment conducted in \cite{Hong2} examined the
changes in entropy of the coordination of isometric force output under
different levels of task demands and feedback from the environment. The goal
of the study was to examine the hypothesis that human motor adaptation can
be characterized as a process of entropy conservation that is reflected in
the compensation of entropy between the task, organism motor output, and
environment. Information entropy of the coordination dynamics relative phase
of the motor output was made conditional on the idealized situation of human
movement, for which the goal was always achieved. Conditional entropy of the
motor output decreased as the error tolerance and feedback frequency were
decreased. Thus, as the likelihood of meeting the task demands was decreased
increased task entropy and/or the amount of information from the environment
is reduced increased environmental entropy, the subjects of this experiment
employed fewer coordination patterns in the force output to achieve the
goal. The conservation of entropy supports the view that context dependent
adaptations in human goal--directed action are guided fundamentally by
natural law and provides a novel means of examining human motor behavior.
This is fundamentally related to the \textit{Heisenberg uncertainty principle%
} \cite{QuLeap} and further supports the argument for the primacy of a
probabilistic approach toward the study of biodynamic cognition systems.\footnote{Our entropic action--amplitude formalism represents a kind of a generalization of the
Haken-Kelso-Bunz (HKB) model of self-organization in the individual's motor
system \cite{HKB,Kelso95}, including: multi-stability, phase
transitions and hysteresis effects, presenting a contrary view to
the purely feedback driven systems. HKB uses the concepts of
synergetics (order parameters, control parameters, instability,
etc) and the mathematical tools of nonlinearly coupled (nonlinear)
dynamical systems to account for self-organized behavior both at
the cooperative, coordinative level and at the level of the
individual coordinating elements. The HKB model stands as a
building block upon which numerous extensions and elaborations
have been constructed. In particular, it has been possible to
derive it from a realistic model of the cortical sheet in which
neural areas undergo a reorganization that is mediated by intra-
and inter-cortical connections. Also, the HKB model describes
phase transitions (`switches') in coordinated human movement as
follows: (i) when the agent begins in the anti-phase mode and
speed of movement is increased, a spontaneous switch to
symmetrical, in-phase movement occurs; (ii) this transition
happens swiftly at a certain critical frequency; (iii) after the
switch has occurred and the movement rate is now decreased the
subject remains in the symmetrical mode, i.e. she does not switch
back; and (iv) no such transitions occur if the subject begins
with symmetrical, in-phase movements. The HKB dynamics of the
order parameter relative phase as is given by a nonlinear
first-order ODE:
$$\dot{\phi} = (\alpha + 2 \beta r^2) \sin\phi - \beta r^2
\sin2\phi,
$$
where $\phi$ is the phase relation (that characterizes the
observed patterns of behavior, changes abruptly at the transition
and is only weakly dependent on parameters outside the phase
transition), $r$ is the oscillator amplitude, while $\alpha,\beta$
are coupling parameters (from which the critical frequency where
the phase transition occurs can be calculated).}

Yet, it is well known that humans possess more
degrees of freedom than are needed to perform any defined motor
task, but are required to co-ordinate them in order to reliably
accomplish high-level goals, while faced with intense motor
variability. In an attempt to explain how this takes place,
Todorov and Jordan have formulated an alternative
theory of human motor co-ordination based on the concept of
stochastic optimal feedback control \cite{Todorov}. They were able to conciliate
the requirement of goal achievement (e.g., grasping an object)
with that of motor variability (biomechanical degrees of freedom).
Moreover, their theory accommodates the idea that the human motor
control mechanism uses internal `functional synergies' to regulate
task--irrelevant (redundant) movement.

Also, a developing field in coordination dynamics involves the theory of
social coordination, which attempts to relate the DC to normal
human development of complex social cues following certain
patterns of interaction. This work is aimed at understanding how
human social interaction is mediated by meta-stability of neural
networks \cite{NeuFuz}. fMRI and EEG are particularly useful in mapping
thalamocortical response to social cues in experimental studies.
In particular, a new theory called the \emph{Phi complex} has been
developed by S. Kelso and collaborators, to provide experimental
results for the theory of social coordination dynamics (see the
recent nonlinear dynamics paper discussing social coordination and
EEG dynamics \cite{Tognoli}). According to this theory, a pair of
phi rhythms, likely generated in the mirror neuron system, is the
hallmark of human social coordination. Using a dual--EEG recording
system, the authors monitored the interactions of eight pairs of
subjects as they moved their fingers with and without a view of
the other individual in the pair.

\section{Formal model of crowd dynamics}

In this section we formally develop a three--step crowd behavior dynamics, conceptualized by transition maps (\ref{id})--(\ref{ad})--(\ref{cd}), in agreement with Haken's synergetics \cite{Haken1,Haken2}.
We first develop a macro--level individual behavior dynamics $\cal ID$. Then we generalize $\cal ID$ into an `orchestrated' behavioral--compositional crowd dynamics $\cal CD$, using a quantum--like micro--level formalism with individual agents representing `crowd quanta'. Finally we develop a meso--level aggregate statistical--field dynamics $\cal AD$, such that composition of the aggregates $\cal AD$ makes--up the crowd.

\subsection{Individual behavior dynamics ($\cal ID$)}

$\cal ID$ transition map (\ref{id}) is developed using the following action--amplitude formalism (see \cite{IA,IAY}):
\begin{enumerate}
\item Macroscopically, as a smooth Riemannian $n-$manifold $M_{\rm ID}$ with steady
force--fields and behavioral paths, modelled by a real--valued classical
{action functional} $S_{\rm ID}[\Phi ]$, of the form
\begin{equation*}
S_{\rm ID}[\Phi ]=\int_{t_{ini}}^{t_{fin}}{L}_{\rm ID}[\Phi ]\,dt,
\end{equation*}%
(where macroscopic paths, fields and geometries are commonly denoted by an
abstract field symbol $\Phi ^{i}$) with the potential--energy based {Lagrangian $L$} given by
\begin{equation*}
{L}_{\rm ID}[\Phi ]=\int d^{n}x\,\mathcal{L}_{\rm ID}(\Phi _{i},\partial _{x^{j}}\Phi
^{i}),
\end{equation*}%
where $\mathcal{L}$ is Lagrangian density, the integral is taken over all $n$ local coordinates $x^{j}=x^{j}(t)$ of the
ID, and $\partial _{x^{j}}\Phi ^{i}$ are time and space partial derivatives
of the $\Phi ^{i}-$variables over coordinates. The standard {least action
principle}
\begin{equation*}
\delta S_{\rm ID}[\Phi ]=0,
\end{equation*}%
gives, in the form of the Euler--Lagrangian equations, a shortest
path, an extreme force--field, with a geometry of minimal
curvature and topology without holes. We will see below that high Riemannian curvature generates chaotic behavior, while holes in the manifold produce topologically induced phase transitions.

\item Microscopically, as a collection of wildly fluctuating and jumping
paths (histories), force--fields and geometries/topologies, modelled by a
complex--valued adaptive path integral, formulated by defining a
multi--phase and multi--path (multi--field and multi--geometry) {transition
amplitude} from the entropy--growing state of Mental~Preparation to the
entropy--conserving state of Physical~Action,
\begin{equation}
\langle \mathrm{Physical~Action\,|\,Mental~Preparation}\rangle_{\mathrm{ID}} :=\int_{%
\mathrm{ID}}\mathcal{D}[\Phi ]\,\mathrm{e}^{iS_{\mathrm{ID}}[\Phi ]}\,
\label{pathInt2}
\end{equation}%
where the functional ID--measure $\mathcal{D}[w\Phi]$ is defined as a weighted product
\begin{equation}
\mathcal{D}[w\Phi]=\lim_{N\to\infty}\prod_{s=1}^{N}w_sd\Phi
_{s}^i, \qquad ({i=1,...,n=con+dis}),\label{prod1}
\end{equation} representing an $\infty-$dimensional neural network
\cite{IA}, with weights $w_s$ updating by the general rule
\begin{equation*}
new\;value(t+1)\;=\;old\;value(t)\;+\;innovation(t).
\end{equation*}
More precisely, the weights $w_s=w_s(t)$ in (\ref{prod1}) are updated according to one of the two
standard neural learning schemes, in which the micro--time level
is traversed in discrete steps, i.e., if $t=t_0,t_1,...,t_s$ then
$t+1=t_1,t_2,...,t_{s+1}$:\footnote{The traditional neural networks approaches are
known for their classes of functions they can
represent. Here we are talking about functions in an
\emph{extensional} rather than merely \emph{intensional} sense;
that is, function can be read as input/output behavior
\cite{Barendregt,Benthem,Forster,Hankin}. This limitation has
been attributed to their low-dimensionality (the largest neural
networks are limited to the order of $10^5$ dimensions
\cite{Izh}). The proposed path integral approach represents a new
family of function-representation methods, which potentially
offers a basis for a fundamentally more expansive solution.}
\begin{enumerate}
    \item A \textit{self--organized}, \textit{unsupervised}
    (e.g., Hebbian--like \cite{Hebb}) learning rule:
\begin{equation}
w_s(t+1)=w_s(t)+ \frac{\sigma}{\eta}(w_s^{d}(t)-w_s^{a}(t)),
\label{Hebb}
\end{equation}
where $\sigma=\sigma(t),\,\eta=\eta(t)$ denote \textit{signal} and
\textit{noise}, respectively, while superscripts $d$ and $a$
denote \textit{desired} and \textit{achieved} micro--states,
respectively; or
    \item A certain form of a \textit{supervised gradient descent
    learning}:
\begin{equation}
w_s(t+1)\,=\,w_s(t)-\eta \nabla J(t), \label{gradient}
\end{equation}
where $\eta $ is a small constant, called the \textit{step size},
or the \textit{learning rate,} and $\nabla J(n)$ denotes the
gradient of the `performance hyper--surface' at the $t-$th
iteration.
\end{enumerate}(Note that we could also use a reward--based, {reinforcement
learning} rule \cite{SB}, in which system learns its {optimal
policy}:
~$
innovation(t)=|reward(t)-penalty(t)|.\,
$)
\end{enumerate}
In this way, we effectively derive a unique and
globally smooth, causal and entropic phase-transition map (\ref{id}),
performed at a macroscopic (global) time--level from some initial time $%
t_{ini}$ to the final time $t_{fin}$. Thus, we have obtained
macro--objects in the ID: a single path described by Newtonian--like
equation of motion, a single force--field described by Maxwellian--like
field equations, and a single obstacle--free Riemannian geometry (with
global topology without holes).

\begin{figure}[htb]
\centering \includegraphics[width=10cm]{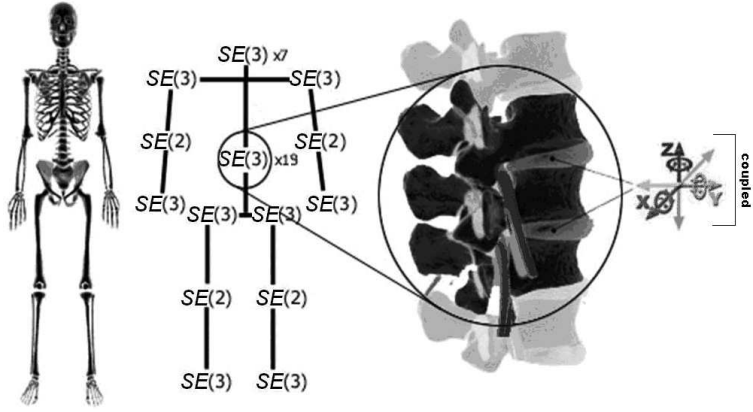}
\caption{\small{Riemannian configuration manifold $M_{\rm ID}$ of human biodynamics is defined
as a topological product $M=\prod_{i}SE(3)^{i}$ of constrained Euclidean $SE(3)$--groups of rigid body motion in 3D Euclidean space (see \cite{GaneshSprBig,GaneshADG}), acting in all major (synovial) human joints. The manifold $M$ is a dynamical structure activated/controlled by potential covariant forces (\ref{cforc1}) produced by a synergetic action of about 640 skeletal muscles \cite{GaneshSprSml}.}}
\label{SpineSE(3)}
\end{figure}

In particular, on the macro--level, we have the ID--paths, that is biodynamical trajectories generated by the Hamilton action principle
$$
\delta S_{\rm ID}[x]=0,
$$
with the {Newtonian action} $S_{\rm ID}[x]$ given by (Einstein's summation convention over repeated indices is always assumed)
\begin{equation}
S_{\rm ID}[x]=\int_{t_{ini}}^{t_{fin}}[\varphi+{1\over2}g_{ij}\,\dot{x}^{i}\dot{x}^{j}]\,dt,
\label{actmot}
\end{equation}
where $\varphi=\varphi(t,x^i)$ denotes the mental LSF--potential field, while the second term,
$$T={1\over2}g_{ij}\,\dot{x}^{i}\dot{x}^{j},$$ represents the physical (biodynamic) kinetic energy
generated by the {Riemannian inertial
metric tensor} $g_{ij}$ of the configuration biodynamic manifold $M_{\rm ID}$ (see Figure \ref{SpineSE(3)}). The corresponding Euler--Lagrangian equations give the
Newtonian equations of human movement
\begin{equation}
\frac{d}{dt}T_{\dot{x}^{i}}-T_{x^{i}}=F_i,
\label{Newton}
\end{equation}
where subscripts denote the partial derivatives and we have defined the covariant muscular forces $F_i=F_i(t,x^i,\dot{x}^{i})$ as negative gradients of the mental potential $\varphi(x^i)$,
\begin{equation}
F_i=-\varphi_{x^i}. \label{cforc1}
\end{equation}
Equation (\ref{Newton}) can be put into
the standard Lagrangian form as
\begin{equation}\frac{d}{dt}L_{\dot{x}^{i}}=L_{x^{i}},\qquad\text{with}\qquad
L=T-\varphi(x^i), \label{LgrEq}
\end{equation}
or (using the Legendre transform) into the forced, dissipative Hamiltonian form \cite{SIAM,GaneshSprSml}
\begin{equation}
\dot{x}^{i} = \partial _{p_{i}}H+\partial _{p_{i}}R, \qquad
\dot{p}_{i} = F_{i}-\partial _{x^{i}}H+\partial _{x^{i}}R, \label{ham1}
\end{equation}
where $p_{i}$ are the generalized momenta (canonically--conjugate to the coordinates $x^i$), $H=H(p,x)$ is the Hamiltonian (total energy function) and $R=R(p,x)$ is the general dissipative function.

The human motor system possesses many independently
controllable components that often allow for more than a
single movement pattern to be performed in order to
achieve a goal. Hence, the motor system is endowed with
a high level of adaptability to different tasks and also
environmental contexts \cite{Hong2}. The multiple SE(3)--dynamics applied to human musculo--skeletal system gives the fundamental law of biodynamics, which is the \emph{covariant force law}:\footnote{This fundamental biodynamic law states that contrary to common perception, acceleration and force are not quantities of
the same nature: while acceleration is a non-inertial vector-field, force is
an inertial co-vector-field. This apparently insignificant difference
becomes crucial in injury prediction/prevention, especially in its derivative form in which the `massless jerk' ($=\dot{a}$) is relatively benign, while the `massive jolt' ($=\dot{F}$) is deadly.}
\begin{equation}
\text{Force co-vector field}=\text{Mass distribution}\times \text{%
Acceleration vector-field}, \label{covForce}
\end{equation}
which is formally written:
\begin{equation*}
F_{i}=g_{ij}a^{j},\qquad (i,j=1,...,n=\dim(M)),
\end{equation*}
where $F_{i}$ are the covariant force/torque components. $g_{ij}$ is the inertial metric tensor of the configuration Riemannian manifold $M=\prod_{i}SE(3)^{i}$ ($g_{ij}$ defines the mass--distribution of the human body),
while $a^{j}$ are the contravariant components of the
linear and angular acceleration vector-field. Both Lagrangian and (topologically equivalent) Hamiltonian development of the covariant force law is fully elaborated in \cite%
{GaneshSprSml,GaneshWSc,GaneshSprBig,GaneshADG}. This is consistent with the postulation that human action is guided primarily by natural law \cite{Kugler}.

On the micro--ID level, instead of each single trajectory defined by the
Newtonian equation of motion (\ref{Newton}), we have an
ensemble of fluctuating and crossing paths on the configuration manifold $M$ with weighted
probabilities (of the unit total sum). This ensemble of
micro--paths is defined by the simplest instance of our adaptive
path integral (\ref{pathInt2}), similar to the Feynman's original
{sum over histories},
\begin{equation}
\langle {\rm Physical~Action\,|\,Mental~Preparation}\rangle_M =\int_{\rm ID}\mathcal{D}[wx]\,
{\mathrm e}^{\mathrm i S[x]}, \label{Feynman}
\end{equation}
where $\mathcal{D}[wx]$ is the functional ID--measure on the
{space of all weighted paths}, and the exponential depends
on the action $S_{\rm ID}[x]$ given by (\ref{actmot}).

\subsection{Crowd behavioral--compositional dynamics ($\cal CD$)}

In this subsection we develop a generic crowd $\cal CD$, as a unique and globally smooth, causal and entropic phase-transition map (\ref{cd}),
in which agents (or, crowd's individual entities) can be both humans and robots. This crowd behavior action takes place in a crowd smooth Riemannian $3n-$manifold $M$. Recall from Figure \ref{SpineSE(3)} that each individual segment of a human body moves in the Euclidean 3--space $\mathbb{R}^3$  according to its own constrained SE(3)--group. Similarly, each individual agent's trajectory, $x^i=x^i(t),~i=1,...n$, is governed by the Euclidean SE(2)--group of rigid body motions in the plane. (Recall
that a Lie group $SE(2)\equiv SO(2)\times \mathbb{R}$ is a set of all $%
3\times 3-$ matrices of the form:
\begin{equation*}
\left[
\begin{array}{ccc}
\cos \theta  & \sin \theta  & x \\
-\sin \theta  & \cos \theta  & y \\
0 & 0 & 1%
\end{array}%
\right] ,
\end{equation*}%
including both rigid translations (i.e., Cartesian $x,y-$coordinates) and
rotation matrix $\left[
\begin{array}{cc}
\cos \theta  & \sin \theta  \\
-\sin \theta  & \cos \theta
\end{array}%
\right] $ in Euclidean plane $\mathbb{R}^{2}$ (see \cite%
{GaneshSprBig,GaneshADG}).)
The crowd configuration manifold $M$ is defined as a union of Euclidean SE(2)--groups for all $n$ individual agents in the crowd, that is
crowd's
configuration $3n-$manifold is defined as a set
\begin{eqnarray}
M &=&\sum_{k=1}^{n}SE(2)^{k}\equiv \sum_{k=1}^{n}SO(2)^{k}\times \mathbb{R}%
^{k},  \label{crwdMan} \\
\text{coordinated by }\mathbf{x}^{k} &\mathbf{=}&\mathbf{\{}%
x^{k},y^{k},\theta ^{k}\},\ (\text{for }k=1,2,...,n).  \notag
\end{eqnarray}
In other words, the crowd configuration manifold $M$ is a \emph{dynamical planar graph} with individual agents' SE(2)--groups of motion in the vertices and time-dependent inter-agent distances $I_{ij} =\left[ x^{i}(t_{i})-x^{j}(t_{j})\right]$ as edges.

Similarly to the individual case, the crowd action functional includes mental cognitive potential and physical kinetic energy, formally given by (with $i,j=1,...,3n$):
\begin{eqnarray}
A[x^i,x^j;t_{i},t_{j}] &=&
\frac{1}{2}\int_{t_{i}}\int_{t_{j}}\,\delta
(I_{ij}^{2})\,\,\dot{x}^{i}(t_{i})\,\dot{x}^{j}(t_{j})\,\,dt_{i}dt_{j}
~+~{\frac{1}{2}}\int_{t}g_{ij}\,\dot{x}^{i}(t)\dot{x}^{j}(t)\,dt,
\label{Fey1} \\
\text{with} && I_{ij}^{2} =\left[ x^{i}(t_{i})-x^{j}(t_{j})\right] ^{2},
\qquad
\text{where \ \ }IN \leq t_{i},t_{j},t\leq OUT.\hspace{2cm}  \notag
\end{eqnarray}
The first term in (\ref{Fey1}) represents the mental potential for the interaction between any two agents $x^i$ and $x^i$ within the total crowd matrix $x^{ij}$. (Although, formally, this term contains cognitive velocities, it still
represents `potential energy' from the physical point of view.) It is defined as a
double integral over a delta function of the square of interval $I^{2}$
between two points on the paths in their individual cognitive LSFs. Interaction
occurs only when this LSF--distance between the two agents $x^i$ and $x^j$ vanishes. Note that the cognitive intentions of any two agents generally occur at different times $t_{i}$ and $t_{j}$
unless $t_{i}=t_{j},$ when cognitive synchronization occurs. This term effectively represents the \emph{crowd cognitive controller} (see \cite{IAY}).

The second term in (\ref{Fey1}) represents {kinetic energy of the physical
interaction of agents}. Namely, after the above cognitive synchronization is completed, the second term of physical kinetic energy is activated in the
common CD manifold, reducing it to just one of the agents' individual manifolds, which is equivalent to the center-of-mass segment in the human musculo-skeletal system. Therefore, from (\ref{Fey1}) we can derive a generic Euler--Lagrangian dynamics that is a composition of (\ref{LgrEq}), which also means that we have in place a generic Hamiltonian dynamics that is a amalgamate of (\ref{ham1}), and the crowd covariant force law (\ref{covForce}), the governing law of crowd biodynamics:
\begin{eqnarray}
\text{Crowd force co-vector field}=\text{Crowd mass distribution}\times \text{%
Crowd acceleration vector-field}, \notag\\
\text{formally:}~~~F_i=g_{ij}a^j,\qquad \text{where}~~g_{ij}~~\text{is the inertial metric tensor of crowd manifold~} M. \label{covForceCrowd}
\end{eqnarray}
The left-hand side of this equation defines forces acting on the crowd, while right-hand defines its mass distribution coupled to the crowd kinematics ($\cal CK$, described in the next subsection).

At the slave level, the adaptive path integral, representing
an $\infty-$dimensional neural network, corresponding to the crowd behavior action (\ref{Fey1}), reads
\begin{equation}
\langle {\rm Physical~Action\,|\,Mental~Preparation}\rangle_{\mathrm{CD}} =\int_{\rm CD}\mathcal{D}[w,x,y]\, {\rm e}^{iA[x,y;t_{i},t_{j}]},  \label{pathInt}
\end{equation}
where the Lebesgue-type integration is performed over all continuous paths $%
x^{i}=x^{i}(t_{i})$ and $y^{j}=y^{j}(t_{j})$, while summation is performed
over all associated discrete Markov fluctuations and jumps. The symbolic differential in the path integral (\ref{pathInt})
represents an {adaptive path measure}, defined as the weighted product
\begin{equation}
\mathcal{D}[w,x,y]=\lim_{N\rightarrow \infty
}\prod_{s=1}^{N}w_{ij}^{s}dx^{i}dy^{j},  \qquad ({i,j=1,...,n}).  \label{prod}
\end{equation}
The quantum--field path integral (\ref{pathInt})--(\ref{prod}) defines the microstate $\cal CD-$level, an
ensemble of fluctuating and crossing paths on the crowd $3n-$manifold $M$.

\subsection{Dissipative crowd kinematics ($\cal CK$)}

The crowd action (\ref{Fey1}) with its amalgamate Lagrangian dynamics (\ref{LgrEq}) and amalgamate Hamiltonian dynamics (\ref{ham1}), as well as the crowd force law (\ref{covForceCrowd})
define the macroscopic crowd dynamics, $\cal CD$. Suppose, for a moment, that $\cal CD$ is force--free and dissipation free, therefore conservative. Now, the basic characteristic of the conservative Lagrangian/Hamiltonian systems evolving in the phase space spanned by the system coordinates and their velocities/momenta, is that their \emph{flow} $\varphi _{t}^L$ (explained below) preserves the phase--space volume. This is proposed by the Liouville theorem, which is the well known fact in
statistical mechanics. However, the preservation of the phase volume causes
structural instability of the conservative system, i.e., the
phase--space spreading effect by which small phase regions $R_{t}$ will
tend to get distorted from the initial one $R_{o}$ during the conservative
system evolution. This problem, governed by entropy growth ($\partial_t S>0$), is much more serious in higher dimensions
than in lower dimensions, since there are so many `directions' in which the
region can locally spread (see \cite{Penrose,GaneshSprBig}). This phenomenon is related to \emph{conservative Hamiltonian chaos} (see section \ref{entr} below).

However, this situation is not very frequent in case of `organized' human crowd. Its self-organization mechanisms are clearly much stronger than the conservative statistical mechanics effects, which we interpret in terms of Prigogine's dissipative structures. Formally, if dissipation of energy in a system is much stronger then its inertial characteristics, then instead of the second-order Newton--Lagrangian dynamic equations of motion, we are actually dealing with the first-order driftless (non-acceleration, non-inertial) kinematic equations of motion, which is related to \emph{dissipative chaos} \cite{Nicolis2}. Briefly, the dissipative crowd flow can be depicted like this: from the set of initial conditions for individual agents, the crowd evolves in time towards the set of the corresponding \emph{entangled attractors,}\footnote{Recall that quantum entanglement is a quantum mechanical phenomenon in which the quantum states of two or more objects are linked together so that one object can no longer be adequately described without full mention of its counterpart -- even though the individual objects may be spatially separated. This interconnection leads to correlations between observable physical properties of remote systems. The related phenomenon of wave-function collapse gives an impression that measurements performed on one system instantaneously influence the other systems entangled with the measured system, even when far apart.

Entanglement has many applications in quantum information theory. Mixed state entanglement can be viewed as a resource for quantum communication.
A common measure of entanglement is the entropy of a mixed quantum state (see, e.g. \cite{QuLeap}). Since a mixed quantum state $\rho$ is a probability distribution over a quantum ensemble, this leads naturally to the definition of the \emph{von Neumann entropy}, ~
$S(\rho) = -  \hbox{Tr} \left( \rho \log_2 {\rho} \right),$~
which is obviously similar to the classical \emph{Shannon entropy} for probability distributions $(p_1, \cdots, p_n)$, defined as
~$S(p_1, \cdots, p_n) = - \sum_i p_i \log_2 p_i.$~
As in statistical mechanics, one can say that the more uncertainty (number of microstates) the system should possess, the larger is its entropy. Entropy gives a tool which can be used to quantify entanglement. If the overall system is pure, the entropy of one subsystem can be used to measure its degree of entanglement with the other subsystems.

The most popular issue in a research on dissipative quantum
brain modelling has been
\textit{quantum entanglement} between the \textit{brain} and its
\textit{environment} \cite{MM,PV}, where the
brain--environment system has an entangled `memory' state,
identified with the ground (vacuum) state $| 0>_{\mathcal{N}}$,
that cannot be factorized into two single--mode
states. (In the Vitiello--Pessa dissipative quantum brain
model \cite{MM,PV}, the evolution of the $\mathcal{N}$--coded
memory system was represented as a trajectory of given initial
condition running over time--dependent states
$|0(t)>_{\mathcal{N}}$, each one minimizing the free energy
functional.) Similar to this microscopic brain--environment
entanglement, we propose a kind of \textit{macroscopic
entanglement} between the operating modes of the \emph{crowd behavior controller} and its biodynamics, which can be considered as a `long--range
correlation'.

Applied externally to the dimension of the crowd $3n-$manifold $M$, entanglement effectively reduces the number of active degrees of freedom in (\ref{crwdMan}).} ~which are mutually separated by fractal (non-integer dimension) separatrices.

In this subsection we elaborate on the dissipative crowd kinematics ($\cal CK$), which is self--controlled and dominates the $\cal CD$ if the crowd's inertial forces are much weaker then the the crowd's dissipation of energy, presented here in the form of nonlinear velocity controllers.

Recall that the essential concept
in dynamical systems theory is the notion of a \textit{vector--field} (that
we will denote by a boldface symbol), which assigns a tangent vector to each
point $p$ in the manifold in case. In particular, $\mathbf{v}$ is a gradient
vector--field if it equals the gradient of some scalar function. A \textit{%
flow--line} of a vector--field $\mathbf{v}$ is a path $\mathbf{\gamma }(t)$
satisfying the vector ODE, ~
$
\mathbf{\dot{\gamma}}(t)=\mathbf{v}(\mathbf{\gamma }(t)),~
$
that is, $\mathbf{v}$ yields the velocity field of the path $\mathbf{\gamma }%
(t)$. The set of all flow lines of a vector--field $\mathbf{v}$ comprises
its flow ~$\varphi _{t}$~ that is (technically, see e.g., \cite%
{GaneshSprBig,GaneshADG}) a one--parameter Lie group of diffeomorphisms
(smooth bijective functions) generated by a vector-field $\mathbf{v}$ on $M$%
, such that
\begin{equation*}
\varphi _{t}\circ \varphi _{s}=\varphi _{t+s},\qquad \varphi _{0}=\text{%
identity},\qquad\text{which gives:}\quad \gamma(t) = \varphi_t(\gamma(0)).
\end{equation*}
Analytically, a vector-field $\mathbf{v}$ is defined as a set of autonomous
ODEs. Its solution gives the flow $\varphi _{t}$, consisting of integral
curves (or, flow lines) $\mathbf{\gamma }(t)$ of the vector--field, such
that all the vectors from the vector-field are tangent to integral curves at
different representative points $p\in M$. In this way, through every
representative point $p\in M$ passes both a curve from the flow and its
tangent vector from the vector-field.
Geometrically, vector-field is defined as a cross-section of the tangent
bundle $TM$ of the manifold $M$.

In general, given an $n$D frame $\{\partial _{i}\}\equiv \{\partial
/\partial x^{i}\}$ on a smooth $n-$manifold $M$ (that is, a basis of tangent
vectors in a local coordinate chart $x^{i}=(x^{1},...,x^{n})\subset M$), we
can define any vector-field $\mathbf{v}$ on $M$ by its components $%
v^{i}=v^{i}(t)$ as
\begin{equation*}
\mathbf{v}=v^{i}\partial _{i}=v^{i}\frac{\partial }{\partial x^{i}}=v^{1}%
\frac{\partial }{\partial x^{1}}+...+v^{n}\frac{\partial }{\partial x^{n}}.
\end{equation*}
Thus, a vector-field $\mathbf{v}\in \mathcal{X}(M)$ (where $\mathcal{X}(M)$
is the set of all smooth vector-fields on $M$) is actually a differential
operator that can be used to differentiate any smooth scalar function $%
f=f(x^{1},...,x^{n})$ on $M$, as a \emph{directional derivative} of $f$ in
the direction of $\mathbf{v}.$ This is denoted simply $\mathbf{v}f$, such
that
\begin{equation*}
\mathbf{v}f=v^{i}\partial _{i}f=v^{i}\frac{\partial f}{\partial x^{i}}=v^{1}%
\frac{\partial f}{\partial x^{1}}+...+v^{n}\frac{\partial f}{\partial x^{n}}.
\end{equation*}

In particular, if $\mathbf{v}=\dot{\gamma}(t)$ is a velocity vector-field of
a space curve $\gamma (t)=(x^{1}(t),...,x^{n}(t)),$ defined by its
components $v^{i}=\dot{x}^{i}(t),$ directional derivative of $f(x^{i})$ in
the direction of $\mathbf{v}$ becomes
\begin{equation*}
\mathbf{v}f=\dot{x}^{i}\partial _{i}f=\frac{dx^{i}}{dt}\frac{\partial f}{%
\partial x^{i}}=\frac{df}{dt}=\dot{f},
\end{equation*}
which is a rate-of-change of $f$ along the curve $\gamma (t)$ at a point $%
x^{i}(t).$

Given two vector-fields, $\mathbf{u}=u^{i}\partial _{i},\mathbf{v}%
=v^{i}\partial _{i}\in \mathcal{X}(M)$, their Lie bracket (or, commutator)
is another vector-field $[\mathbf{u},\mathbf{v}]$ $\in \mathcal{X}(M),$
defined by
\begin{equation*}
\lbrack \mathbf{u},\mathbf{v}]=\mathbf{uv}-\mathbf{vu}=u^{i}\partial
_{i}\,v^{j}\partial _{j}-v^{j}\partial _{j}\,u^{i}\partial _{i},
\end{equation*}
which, applied to any smooth function $f$ on $M,$ gives
\begin{equation*}
\lbrack \mathbf{u},\mathbf{v}](f)=\mathbf{u}\left( \mathbf{v}(f)\right) -%
\mathbf{v}\left( \mathbf{u}(f)\right) .
\end{equation*}

The Lie bracket measures the failure of `mixed directional derivatives' to
commute. Clearly, mixed partial derivatives \textit{do} commute, ~$\lbrack \partial _{i},\partial _{j}]=0$, while in general it is \emph{not} the case, ~$\lbrack \mathbf{u},\mathbf{v}]\neq 0$. In addition, suppose that $\mathbf{u}$ generates the flow $\varphi _{t}$ and $\mathbf{v}$
generates the flow $\varphi _{s}$. Then, for any smooth function $f$ on $M,$
we have at any point $p$ on $M,$%
\begin{equation*}
\lbrack \mathbf{u},\mathbf{v}](f)(p)=\frac{\partial ^{2}}{\partial t\partial
s}\left( f(\varphi _{s}(\varphi _{t}(p))\right) -f(\varphi _{t}(\varphi
_{s}(p))),
\end{equation*}%
which means that in $f(\varphi _{s}(\varphi _{t}(p))$ we are starting at $p$%
, flowing along $\mathbf{v}$ a little bit, then along $\mathbf{u}$ a little
bit, and then evaluating $f$, while in $f(\varphi _{t}(\varphi _{s}(p))$ we
are flowing first along $\mathbf{u}$ and then $\mathbf{v}$. Therefore, the
Lie bracket infinitesimally measures how these flows fail to commute.

The Lie bracket satisfies the following three properties (for any three
vector-fields $\mathbf{u,v,w}\in M$ and two constants $a,b$ -- thus forming a Lie algebra on the crowd manifold $M$):

(i)~~ $[\mathbf{u},\mathbf{v}]=-[\mathbf{v},\mathbf{u}]-$ skew-symmetry;

(ii)~ $[\mathbf{u},a\mathbf{v}+b\mathbf{w}]=a[\mathbf{u},\mathbf{v}]+b[%
\mathbf{u},\mathbf{w}]-$ bilinearity; ~and

(iii) $[\mathbf{u},[\mathbf{v},\mathbf{w}]]+[\mathbf{v},[\mathbf{w},\mathbf{u%
}]]+[\mathbf{w},[\mathbf{u},\mathbf{v}]]-$ Jacobi identity.\newline
A new set of vector-fields on $M$ can be generated by repeated Lie brackets
of $\mathbf{u,v,w}\in M$.

The Lie bracket is a standard tool in geometric nonlinear control theory
(see, e.g. \cite{GaneshSprBig,GaneshADG}). Its action on vector-fields can
be best visualized using the popular car parking example, in which the
driver has two different vector--field transformations at his disposal. They
can turn the steering wheel, or they can drive the car forward or backward.
Here, we specify the state of a car by four coordinates: the $(x,y)$
coordinates of the center of the rear axle, the direction $\theta $ of the
car, and the angle $\phi $ between the front wheels and the direction of the
car. $l$ is the constant length of the car. Therefore, the 4D configuration
manifold of a car is a set $M\equiv SO(2)\times \mathbb{R}^{2},$ coordinated
by $\mathbf{x=\{}x,y,\theta ,\phi \}$, which is slightly more complicated
than the individual crowd agent's 3D configuration manifold $SE(2)\equiv
SO(2)\times \mathbb{R},$ coordinated by $\mathbf{x=\{}x,y,\theta \}$. The
driftless car kinematics can be defined as a vector ODE:
\begin{equation}
\mathbf{\,\dot{x}}=\mathbf{u}(\mathbf{x})\,c_{1}+\mathbf{v}(\mathbf{x}%
)\,c_{2},  \label{Car2}
\end{equation}%
with two vector--fields, $\mathbf{u},\mathbf{v}\in \mathcal{X}(M)$, and two
scalar control inputs, $c_{1}$ and $c_{2}$. The infinitesimal car--parking transformations will be the following vector--fields
\begin{eqnarray*}
\mathbf{u}(\mathbf{x}) &\equiv &\text{\textsc{drive}}=\cos \theta \frac{%
\partial }{\partial x}+\sin \theta \frac{\partial }{\partial y}+\frac{\tan
\phi }{l}\frac{\partial }{\partial \theta }\equiv \left(
\begin{array}{c}
\cos \theta \\
\sin \theta \\
\frac{1}{l}\tan \phi \\
0%
\end{array}%
\right) , \\
\text{and}\qquad \mathbf{v}(\mathbf{x}) &\equiv &\text{\textsc{steer}}=\frac{%
\partial }{\partial \phi }\equiv \left(
\begin{array}{c}
0 \\
0 \\
0 \\
1%
\end{array}%
\right) .
\end{eqnarray*}%
The car kinematics (\ref{Car2}) therefore expands into a matrix ODE:
\begin{equation*}
\left(
\begin{array}{c}
\dot{x} \\
\dot{y} \\
\dot{\theta} \\
\dot{\phi}%
\end{array}%
\right) =\text{\textsc{drive}}\cdot c_{1\text{ }}+\text{\textsc{steer}}\cdot
c_{2\text{ }}\equiv \left(
\begin{array}{c}
\cos \theta \\
\sin \theta \\
\frac{1}{l}\tan \phi \\
0%
\end{array}%
\right) \cdot c_{1\text{ }}+\left(
\begin{array}{c}
0 \\
0 \\
0 \\
1%
\end{array}%
\right) \cdot c_{2\text{ }}.
\end{equation*}

However, \textsc{steer}\ and \textsc{drive}\ do not commute (otherwise we
could do all your steering at home before driving of on a trip). Their
combination is given by the Lie bracket
\begin{equation*}
\lbrack \mathbf{v},\mathbf{u}]\equiv \lbrack \text{\textsc{steer}},\text{%
\textsc{drive}}]=\frac{1}{l\cos ^{2}\phi }\frac{\partial }{\partial \theta }%
\equiv \text{\textsc{wriggle}}.
\end{equation*}%
The operation $[\mathbf{v},\mathbf{u}]\equiv $ \textsc{wriggle }$\equiv
\lbrack $\textsc{steer}$,$\textsc{drive}$]$ is the infinitesimal version of
the sequence of transformations: steer, drive, steer back, and drive back,
i.e.,
\begin{equation*}
\{\text{\textsc{steer}},\text{\textsc{drive}},\text{\textsc{steer}}^{-1},%
\text{\textsc{drive}}^{-1}\}.
\end{equation*}%
Now, \textsc{wriggle}\ can get us out of some parking spaces, but not tight
ones: we may not have enough room to \textsc{wriggle} out. The usual tight
parking space restricts the \textsc{drive}\ transformation, but not \textsc{%
steer}. A truly tight parking space restricts \textsc{steer}\ as well by
putting your front wheels against the curb.

Fortunately, there is still another commutator available:
\begin{eqnarray*}
\lbrack \mathbf{u},[\mathbf{v},\mathbf{u}]] &\equiv &[\text{\textsc{drive}},[%
\text{\textsc{steer}},\text{\textsc{drive}}]]=[[\mathbf{u},\mathbf{v}],%
\mathbf{u}]\equiv \\
\lbrack \text{\textsc{drive}},\text{\textsc{wriggle}}] &=&\frac{1}{l\cos
^{2}\phi }\left( \sin \theta \frac{\partial }{\partial x}-\cos \theta \frac{%
\partial }{\partial y}\right) \equiv \text{\textsc{slide}}.
\end{eqnarray*}%
The operation $[[\mathbf{u},\mathbf{v}],\mathbf{u}]\equiv $ \textsc{slide }$%
\equiv \lbrack $\textsc{drive}$,$\textsc{wriggle}$]$ is a displacement at
right angles to the car, and can get us out of any parking place. We just
need to remember to steer, drive, steer back, drive some more, steer, drive
back, steer back, and drive back:
\begin{equation*}
\{\text{\textsc{steer}},\text{\textsc{drive}},\text{\textsc{steer}}^{-1},%
\text{\textsc{drive}},\text{\textsc{steer}},\text{\textsc{drive}}^{-1},\text{%
\textsc{steer}}^{-1},\text{\textsc{drive}}^{-1}\}.
\end{equation*}%
We have to reverse steer in the middle of the parking place. This is not
intuitive, and no doubt is part of a common problem with parallel parking.

Thus, from only two controls, $c_{1}$ and $c_{2}$, we can form the
vector--fields \textsc{drive }$\equiv \mathbf{u}$, \textsc{steer }$\equiv
\mathbf{v}$, \textsc{wriggle }$\equiv $ $[\mathbf{v},\mathbf{u}],$ and
\textsc{slide }$\equiv \lbrack \lbrack \mathbf{u},\mathbf{v}],\mathbf{u}]$,
allowing us to move anywhere in the car configuration manifold $M\equiv
SO(2)\times \mathbb{R}^{2}$. All above computations are straightforward in $Mathematica^{TM}$\footnote{The above computations could instead be done in other available packages, such as Maple, by suitably translating the provided example code.} if we define the following three symbolic functions:

1. Jacobian matrix: ~~JacMat[v\_List, x\_List] := Outer[D, v, x];

2. Lie bracket: ~~LieBrc[u\_List, v\_List, x\_List] := JacMat[v, x] . u - JacMat[u, x] . v;

3. Repeated Lie bracket: ~~Adj[u\_List, v\_List, x\_List, k\_] := \\
$~$\hspace{5.6cm} If[k == 0, v, LieBrc[u, Adj[u, v, x, k - 1], x] ];

In case of the human crowd, we have a slightly simpler, but multiplied problem, i.e., superposition of $n$ individual agents' motions. So, we can define the dissipative crowd kinematics as a system of $n$ vector ODEs:
\begin{equation}
\,\mathbf{\dot{x}}^{k}=\mathbf{u}^{k}(\mathbf{x})\,c_{1}^{k}+\mathbf{v}^{k}(%
\mathbf{x})\,c_{2}^{k},\qquad\text{where}  \label{CrwdKin1}
\end{equation}
\begin{eqnarray*}
\mathbf{u}^{k}(\mathbf{x}) &\equiv &\text{\textsc{drive}}^{k}=\cos
^{k}\theta \frac{\partial }{\partial x^{k}}+\sin ^{k}\theta \frac{\partial }{%
\partial y^{k}}\equiv \left(
\begin{array}{c}
\cos ^{k}\theta  \\
\sin ^{k}\theta  \\
0%
\end{array}%
\right) ,\qquad \text{and} \\
\mathbf{v}^{k}(\mathbf{x}) &\equiv &\text{\textsc{steer}}^{k}=\frac{\partial
}{\partial \theta ^{k}}\equiv \left(
\begin{array}{c}
0 \\
0 \\
1%
\end{array}%
\right) ,\qquad \text{while \ }c_{1}^{k}\ \text{\ and \ }c_{2}^{k}\text{ \
are crowd controls.}
\end{eqnarray*}%
Thus, the crowd kinematics (\ref{CrwdKin1}) expands into the matrix ODE:
\begin{equation}
\left(
\begin{array}{c}
\dot{x} \\
\dot{y} \\
\dot{\theta}%
\end{array}%
\right) =\text{\textsc{drive}}^{k}\cdot c_{1\text{ }}^{k}+\text{\textsc{steer%
}}^{k}\cdot c_{2\text{ }}^{k}\equiv \left(
\begin{array}{c}
\cos ^{k}\theta  \\
\sin ^{k}\theta  \\
0%
\end{array}%
\right) \cdot c_{1\text{ }}^{k}+\left(
\begin{array}{c}
0 \\
0 \\
1%
\end{array}%
\right) \cdot c_{2\text{ }}^{k}.  \label{CrwdKin2}
\end{equation}

\begin{figure}[htb]
\centering \includegraphics[width=14cm]{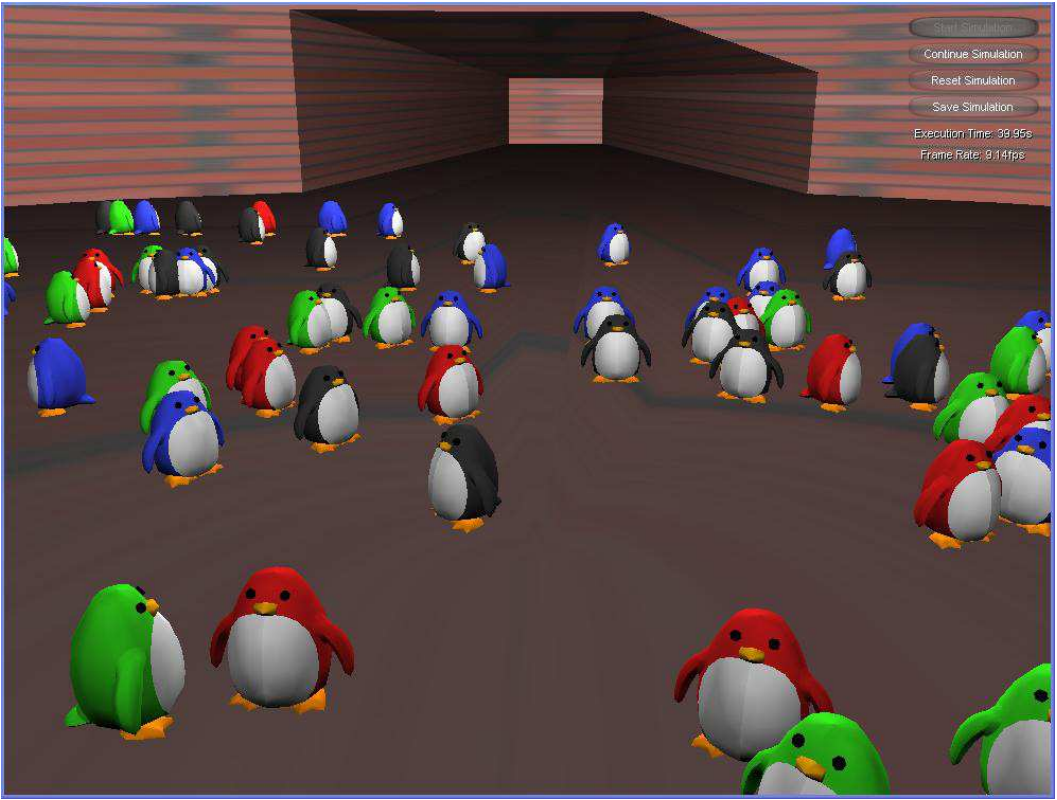}
\caption{Driving and steering random $SE(2)-$dynamics of 120 penguin-like robots (with embedded collision-detection). Compare with \cite{Arizona}.}
\label{IvCrowd1}
\end{figure}

A 3D simulation of random, dissipative crowd kinematics (\ref{CrwdKin1})--(\ref{CrwdKin2}) of 120 penguin-like $SE(2)-$robots, developed in C++/DirX is presented in Figure \ref{IvCrowd1}.

The dissipative crowd kinematics (\ref{CrwdKin1})--(\ref{CrwdKin2}) obeys the set of $n$-tuple integral rules of motion that are similar (though slightly simpler) to the above rules of the car kinematics, including the following derived vector-fields:

\textsc{wriggle}$^{k}$\textsc{\ }$\equiv \lbrack $\textsc{steer}$^{k},$%
\textsc{drive}$^{k}]\text{\textsc{\ }}\equiv \lbrack \mathbf{v}^{k},\mathbf{u%
}^{k}]$ \  and  \
\textsc{slide}$^{k}$\textsc{\ }$\equiv \lbrack $\textsc{drive}$^{k},$\textsc{%
wriggle}$^{k}]\equiv \lbrack \lbrack \mathbf{u}^{k},\mathbf{v}^{k}],\mathbf{u%
}^{k}].$

Thus, controlled by the two vector controls $c_{1}^{k}$ and $c_{2}^{k},$ the
crowd can form the vector--fields: \textsc{drive }$\equiv \mathbf{u}^{k}$,
\textsc{steer }$\equiv \mathbf{v}^{k}$, \textsc{wriggle }$\equiv $ $[\mathbf{%
v}^{k},\mathbf{u}^{k}],$ and \textsc{slide }$\equiv \lbrack \lbrack \mathbf{u%
}^{k},\mathbf{v}^{k}],\mathbf{u}^{k}]$, allowing it to move anywhere within
its configuration manifold $M$ given by (\ref{crwdMan}). Solution of the dissipative crowd kinematics (\ref{CrwdKin1})--(\ref{CrwdKin2}) defines the dissipative crowd flow, $\phi_t^{K}$.

Now, the general $\cal CD$--$\cal CK$ crowd behavior can be defined as a amalgamate flow (behavior--Lagrangian flow, $\phi_t^{L}$, plus dissipative kinematic flow, $\phi_t^{K}$) on the crowd manifold $M$ defined by (\ref{crwdMan}), \begin{equation*}C_t=\phi_t^{L}+\phi_t^{K}:t\mapsto (M(t),g(t)),\end{equation*} which is
a one-parameter family of homeomorphic (topologically equivalent) Riemannian manifolds\footnote{Proper differentiation of vector and tensor fields on a smooth Riemannian
manifold (like the crowd $3n-$manifold $M$) is performed using the \textit{%
Levi--Civita covariant derivative} (see, e.g., \cite{GaneshSprBig,GaneshADG}%
). Formally, let $M$ be a Riemannian $N-$manifold with the tangent bundle $TM
$ and a local coordinate system $\{x^{i}\}_{i=1}^{N}$ defined in an open set
$U\subset M$. The covariant derivative operator, $\nabla _{X}:C^{\infty
}(TM)\rightarrow C^{\infty }(TM)$, is the unique linear map such that for
any vector-fields $X,Y,Z,$ constant $c$, and scalar function $f$ the
following properties are valid:
\[
\nabla _{X+cY}=\nabla _{X}+c\nabla _{Y},\qquad \nabla _{X}(Y+fZ)=\nabla
_{X}Y+(Xf)Z+f\nabla _{X}Z,\qquad \nabla _{X}Y-\nabla _{Y}X=[X,Y],
\]%
where $[X,Y]$ is the Lie bracket of $X$ and $Y$. In local coordinates, the
metric $g$ is defined for any orthonormal basis $(\partial _{i}=\partial
/\partial x^{i})$ in $U\subset M$ by \ $g_{ij}=g(\partial _{i},\partial
_{j})=\delta _{ij},$ $\ \partial _{k}g_{ij}=0.$ Then the affine \textit{%
Levi--Civita connection} is defined on $M$ by
\[
\nabla _{\partial _{i}}\partial _{j}=\Gamma _{ij}^{k}\partial _{k},~~\text{ \
\ where \ }~~\Gamma _{ij}^{k}=\frac{1}{2}g^{kl}\left( \partial
_{i}g_{jl}+\partial _{j}g_{il}-\partial _{l}g_{ij}\right) \text{ \ are the
Christoffel symbols}.
\]

Now, using the covariant derivative operator $\nabla _{X}$ we can define the
\textit{Riemann curvature} $(3,1)-$tensor $\mathfrak{Rm}$ by
\[
\mathfrak{Rm}(X,Y)Z=\nabla _{X}\nabla _{Y}Z-\nabla _{Y}\nabla _{X}Z-\nabla
_{\lbrack X,Y]}Z,
\]%
which measures the curvature of the manifold by expressing how
noncommutative covariant differentiation is. The $(3,1)-$components $%
R_{ijk}^{l}$ of $\mathfrak{Rm}$ are defined in $U\subset M$ by
\[
\mathfrak{Rm}\left( \partial _{i},\partial _{j}\right) \partial
_{k}=R_{ijk}^{l}\partial _{l},~~\text{ \ \ or \ }~~~R_{ijk}^{l}=\partial
_{i}\Gamma _{jk}^{l}-\partial _{j}\Gamma _{ik}^{l}+\Gamma _{jk}^{m}\Gamma
_{im}^{l}-\Gamma _{ik}^{m}\Gamma _{jm}^{l}.
\]%
Also, the Riemann $(4,0)-$tensor $R_{ijkl}=g_{lm}R_{ijk}^{m}$ is defined as
the $g-$based inner product on $M$,
\[
R_{ijkl}=\left\langle \mathfrak{Rm}\left( \partial _{i},\partial _{j}\right)
\partial _{k},\partial _{l}\right\rangle .
\]

The first and second Bianchi identities for the Riemann $(4,0)-$tensor $%
R_{ijkl}$ hold,
\[
R_{ijkl}+R_{jkil}+R_{kijl}=0,\qquad \nabla _{i}R_{jklm}+\nabla
_{j}R_{kilm}+\nabla _{k}R_{ijlm}=0,
\]%
while the twice contracted second Bianchi identity reads: \ $2\nabla
_{j}R_{ij}=\nabla _{i}R.$

The $(0,2)$ \textit{Ricci tensor} $\mathfrak{Rc}$ is the trace of the
Riemann $(3,1)-$tensor $\mathfrak{Rm}$,
\[
\mathfrak{Rc}(Y,Z)+\mathrm{tr}(X\rightarrow \mathfrak{Rm}(X,Y)Z),~~\text{ \ \
so that \ }~~\mathfrak{Rc}(X,Y)=g(\mathfrak{Rm}(\partial _{i},X)\partial
_{i},Y),
\]%
Its components $R_{jk}=\mathfrak{Rc}\left( \partial _{j},\partial
_{k}\right) $ are given in $U\subset M$ by the contraction
\[
R_{jk}=R_{ijk}^{i},~\quad \text{or \ }~~R_{jk}=\partial _{i}\Gamma
_{jk}^{i}-\partial _{k}\Gamma _{ji}^{i}+\Gamma _{mi}^{i}\Gamma
_{jk}^{m}-\Gamma _{mk}^{i}\Gamma _{ji}^{m}.
\]

Finally, the scalar curvature $R$ is the trace of the Ricci tensor $%
\mathfrak{Rc}$, given in $U\subset M$ by:~ $R=g^{ij}R_{ij}. $} $(M,g=g_{ij})$, parameterized by a `time' parameter $t$. That is, $C_{t}$ can
be used for describing smooth deformations of the crowd manifold $M$ over time. The manifold family $(M(t),g(t))$ at time $t$ determines the manifold family $(M(t+dt),g(t+dt))
$ at an infinitesimal time $t+dt$ into the future, according to some presecribed geometric flow, like the celebrated \emph{Ricci
flow} \cite{Ham82,4-manifold,Harnack,surface} (that was an instrument for a proof of a 100--year old Poincar\'e conjecture),
\begin{equation}
\partial _{t}g_{ij}(t)=-2R_{ij}(t),  \label{RF}
\end{equation}%
where $R_{ij}$ is the Ricci curvature tensor of the crowd manifold $M$ and $\partial _{t}g(t)$ is defined as
\begin{equation}
\partial _{t}g(t)\equiv \frac{d}{dt}g(t):=\lim_{dt\rightarrow 0}\frac{%
g(t+dt)-g(t)}{dt}. \label{RFlim}
\end{equation}%

\subsection{Aggregate behavioral--compositional dynamics ($\mathcal{AD}$)} \label{aggreg}

To formally develop the meso-level aggregate behavioral--compositional dynamics ($\mathcal{AD}$), we start with the crowd path integral (\ref{pathInt}), which can
be redefined if we Wick--rotate
the time variable $t$ to imaginary values, $t\mapsto \tau={\mathrm
i} t$, thereby transforming the Lorentzian path integral in real time into the Euclidean path integral in imaginary time. Furthermore, if we rectify the time axis back to the real line, we get the adaptive SFT--partition function as our proposed $\mathcal{AD}$--model:
\begin{equation}
\langle {\rm Physical~Action\,|\,Mental~Preparation}\rangle_{\mathrm{AD}} =\int_{\rm CD}\mathcal{D}[w,x,y]\, {\rm e}^{-A[x,y;t_{i},t_{j}]}. \label{sft}
\end{equation}

The adaptive $\mathcal{AD}$--transition amplitude $\langle {\rm Physical~Action\,|\,Mental~Preparation}\rangle_{\mathrm{AD}}$ as defined by the SFT--partition function (\ref{sft}) is a general model of a\textit{Markov stochastic process}. Recall that Markov process is a random
process characterized by a \emph{lack of memory}, i.e., the
statistical properties of the immediate future are uniquely
determined by the present, regardless of the past (see, e.g. \cite{Gardiner,GaneshSprBig}).
The $N-$dimensional Markov process can be defined by the Ito
stochastic differential equation,
\begin{eqnarray}
dx_{i}(t) &=&A_{i}[x^{i}(t),t]dt+B_{ij}[x^{i}(t),t]\,dW^{j}(t), \\
x^{i}(0) &=&x_{i0},\qquad (i,j=1,\dots ,N) \label{Ito1}
\end{eqnarray}%
or corresponding \emph{Ito stochastic integral equation}
\begin{eqnarray}
x^{i}(t)=x^{i}(0)+\int_{0}^{t}ds\,A_{i}[x^{i}(s),s]+\int_{0}^{t}dW^{j}(s)%
\,B_{ij}[x^{i}(s),s], \label{Ito2}
\end{eqnarray}%
in which $x^{i}(t)$ is the variable of interest, the vector
$A_{i}[x(t),t]$ denotes deterministic \emph{drift}, the matrix
$B_{ij}[x(t),t]$ represents
continuous stochastic \emph{diffusion fluctuations}, and $W^{j}(t)$ is an $%
N-$ variable \textit{Wiener process} (i.e., generalized Brownian
motion \cite{Gardiner}) and $$dW^{j}(t)=W^{j}(t+dt)-W^{j}(t).$$

The two Ito equations (\ref{Ito1})--(\ref{Ito2}) are equivalent to the general \textit{Chapman--Kolmogorov probability equation} (see equation (\ref{CK}) below). There are
three well known special cases of the
Chapman--Kolmogorov equation (see \cite{Gardiner}):
\begin{enumerate}
    \item When both $B_{ij}[x(t),t]$ and $W(t)$ are zero, i.e., in the case of
pure deterministic motion, it reduces to the \textit{Liouville
equation}
\begin{eqnarray*}
\partial _{t}P(x',t'|x'',t'')=-\sum_{i}\frac{\partial }
{\partial x^{i}}\left\{ A_{i}[x(t),t]\,P(x'%
,t'|x'',t'')\right\} .
\end{eqnarray*}
    \item When only $W(t)$ is zero, it reduces to the
    \textit{Fokker--Planck equation}
\begin{eqnarray*}
\partial _{t}P(x',t'|x'',t'')&=&
-\sum_{i}\frac{\partial }{\partial x^{i}}\left\{ A_{i}[x(t),t]\,P(x'%
,t'|x'',t'')\right\} \\
&+&\frac{1}{2}\sum_{ij}\frac{\partial ^{2}}{\partial x^{i}\partial
x^{j}}\left\{ B_{ij}[x(t),t]\,P(x',t'|x'',t'')\right\} .
\end{eqnarray*}
    \item When both $A_{i}[x(t),t]$ and $B_{ij}[x(t),t]$ are zero,
i.e., the state--space consists of integers only, it reduces to
the \textit{Master equation} of discontinuous jumps
\begin{eqnarray*}
\partial _{t}P(x',t'|x'',t'') =
\int dx\,W(x'|x'',t)\,P(x',t'|x'',t'')-\int dx\,W(x''|x',t)\,P(%
x',t'|x'',t'').
\end{eqnarray*}
\end{enumerate}

The \textit{Markov assumption} can now be formulated in terms of
the conditional probabilities $P(x^{i},t_{i})$: if the times
$t_{i}$ increase from right to left, the conditional probability
is determined entirely by the knowledge of the most recent
condition. Markov process is generated by a set of conditional
probabilities whose probability--density $P=P(x',t'|x'',t'')$
evolution obeys the general \textit{Chapman--Kolmogorov
integro--differential equation}
\begin{eqnarray*}
\partial _{t}P &=&-\sum_{i}\frac{\partial }{\partial x^{i}}\left\{
A_{i}[x(t),t]\,P\right\}
~+~\frac{1}{2}\sum_{ij}\frac{\partial ^{2}}{\partial x^{i}\partial x^{j}}%
\left\{ B_{ij}[x(t),t]\,P\right\}\\
&+&\int dx\left\{ W(x^{\prime
}|x^{\prime \prime },t)\,P-W(x^{\prime \prime }|x^{\prime
},t)\,P\right\} \label{CK}
\end{eqnarray*}
including \emph{deterministic drift}, \emph{diffusion
fluctuations} and \emph{discontinuous jumps} (given respectively
in the first, second and third terms on the r.h.s.). This general Chapman--Kolmogorov integro-differential
equation (\ref{CK}), with its conditional probability density evolution,
$P=P(x',t'|x'',t'')$, is represented by our SFT--partition function (\ref{sft}).

Furthermore, discretization of the adaptive SFT--partition function (\ref{sft}) gives the standard
\emph{partition function}
\begin{equation}
Z=\sum_j{\mathrm e}^{-w_jE^j/T}, \label{partition}
\end{equation}
where $E^j$ is the motion energy eigenvalue (reflecting each
possible motivational energetic state), $T$ is the
temperature--like environmental control parameter, and the sum
runs over all ID energy eigenstates (labelled by the index
$j$). From (\ref{partition}), we can calculate the \emph{transition entropy}, as $S = k_B\ln Z$ (see the next section).

\section{Entropy, chaos and phase transitions in the crowd\\ manifold} \label{entr}

Recall that nonequilibrium phase transitions \cite%
{Haken1,Haken2,Haken3,Haken4,HakenBrain} are phenomena which bring about {qualitative}
physical changes at the macroscopic level in presence of the same
microscopic forces acting among the constituents of a system. In this
section we extend the $\cal CD$ formalism to incorporate both algorithmic and geometrical entropy as well as dynamical chaos \cite%
{Ott90,TacaNODY,StrAttr,Complexity} between the entropy--growing phase of Mental Preparation and the entropy--conserving phase of Physical Action, together with the associated topological phase transitions.

\subsection{Algorithmic entropy}

The Boltzmann and Shannon (hence also Gibbs entropy, which is Shannon
entropy scaled by $k\ln2$, where $k$ is the Bolzmann constant) entropy
definitions involve the notion of \emph{ensembles}. Membership of
microscopic states in ensembles defines the probability density function
that underpins the entropy function; the result is that the entropy
of a definite and completely known microscopic state is precisely
zero. Bolzmann entropy defines the probabilistic model of the system
by effectively discarding part of the information about the system,
while the Shannon entropy is concerned with measuring the ignorance
of the observer -- the amount of missing information -- about the system.

Zurek proposed a new physical entropy measure that can be applied
to individual microscopic system states and does not use the ensemble
structure. This is based on the notion of a fixed individually random
object provided by Algorithmic Information Theory and Kolmogorov Complexity:
put simply, the randomness $K(x)$ of a binary string $x$ is the
length in terms of number of bits of the smallest program $p$ on
a universal computer that can produce $x$.

While this is the basic idea, there are some important technical details
involved with this definition. The randomness definition uses the
prefix complexity $K(.)$ rather than the older Kolmogorov complexity
measure $C(.)$: the prefix complexity $K(x|y)$ of $x$ given $y$
is the Kolmogorov complexity $C_{\phi_{u}}(x|y)=\min\left\{ p\:|\: x=\phi_{u}(\left\langle y,p\right\rangle )\right\} $
(with the convention that $C_{\phi_{u}}(x|y)=\infty$ if there is
no such $p$) that is taken with respect to a reference universal
partial recursive function $\phi_{u}$ that is a universal prefix
function. Then the prefix complexity $K(x)$ of $x$ is just $K(x|\varepsilon)$
where $\varepsilon$ is the empty string. A partial recursive prefix
function $\phi:M\rightarrow\mathbb{N}$ is a partial recursive
function such that if $\phi(p)<\infty$ and $\phi(q)<\infty$ then
$p$ is not a proper prefix of $q$: that is, we restrict the complexity
definition to a set of strings (which are descriptions of effective
procedures) such that none is a proper prefix of any other. In this
way, all effective procedure descriptions are \emph{self-delimiting}:
the total length of the description is given within the description
itself. A universal prefix function $\phi_{u}$is a prefix function
such that $\forall n\in\mathbb{N\;}\phi_{u}(\left\langle y,\left\langle n,p\right\rangle \right\rangle )=\phi_{n}(\left\langle y,p\right\rangle )$,
where $\phi_{n}$ is numbered $n$ according to some Godel numbering
of the partial recursive functions; that is, a universal prefix function
is a partial recursive function that simulates any partial recursive
function. Here, $\left\langle x,y\right\rangle $ stands for a total
recusive one-one mapping from $\mathbb{N}$$\times\mathbb{N}$ into
$\mathbb{N}$, $\left\langle x_{1},x_{2},\ldots,x_{n}\right\rangle =\left\langle x_{1},\left\langle x_{2},\ldots,x_{n}\right\rangle \right\rangle $,
$\mathbb{N}$ is the set of natural numbers, and $M=\left\{ 0,1\right\} ^{*}$is
the set of all binary strings.

This notion of entropy circumvents the use of probability to give
a concept of entropy that can be applied to a fully specified macroscopic
state: the algorithmic randomness of the state is the length of the
shortest possible effective description of it. To illustrate, suppose
for the moment that the set of microscopic states is countably infinite,
with each state identified with some natural number. It is known that
the discrete version of the Gibbs entropy (and hence of Shannon's
entropy) and the algorithmic entropy are asymptotically consistent
under mild assumptions. Consider a system with a countably infinite
set of microscopic states $X$ supporting a probability density function
$P(.)$ so that $P(x)$ is the probability that the system is in microscopic
state $x\in X$. Then the Gibbs entropy is $S_{G}(P)=-(k\ln2)\underset{x\in X}{\sum}P(x)\log P(x)$
(which is Shannon's information-theoretic entropy $H(P)$ scaled by
$k\ln2$). Supposing that $P(.)$ is recursive, then $S_{G}(P)=(k\ln2)\underset{x\in X}{\sum}P(x)K(x)+C$,
where $C_{\phi}$ is a constant depending only on the choice of the
reference universal prefix function $\phi$. Hence, as a measure of
entropy, the function $K(.)$ manifests the same kind of behavior
as Shannon's and Gibbs entropy measures.

Zurek's proposal was of a new physical entropy measure that includes
contributions from both the randomness of a state and ignorance about
it. Assume now that we have determined the macroscopic parameters
of the system, and encode this as a string - which can always be converted
into an equivalent binary string, which is just a natural number under
a standard encoding. It is standard to denote the binary string and
its corresponding natural number interchangeably; here let $x$ be
the encoded macroscopic parameters. Zurek's definition of \emph{algorithmic
entropy} of the macroscopic state is then $K(x)+H_{x}$, where $H_{x}=S_{B}(x)/(k\ln2)$,
where $S_{B}(x)$ is the Bolzmann entropy of the system constrained
by $x$ and $k$ is Bolzmann's constant; the physical version of the
algorithmic entropy is therefore defined as $S_{A}(x)=(k\ln2)(K(x)+H_{x})$.
Here $H_{x}$represents the level of ignorance about the microscopic
state, given the parameter set $x$; it can decrease towards zero
as knowledge about the state of the system increases, at which point
the algorithmic entropy reduces to the Bolzmann entropy.

\subsection{Ricci flow and Perelman entropy--action on the crowd manifold}

Recall that the inertial metric crowd flow, $C_{t}:t\mapsto (M(t),g(t))$ on the crowd $3n-$manifold (\ref{crwdMan}) is
a one-parameter family of homeomorphic Riemannian manifolds $(M,g)$, evolving by the Ricci flow (\ref{RF})--(\ref{RFlim}).

Now, given a smooth scalar function $u:M \rightarrow \mathbb{R}$ on the Riemannian crowd $3n-$manifold $M $, its Laplacian operator $\Delta $ is
locally defined as
\begin{equation*}
\Delta u=g^{ij}\nabla _{i}\nabla _{j}u,
\end{equation*}%
where $\nabla _{i}$ is the covariant derivative (or, Levi--Civita
connection). We say that a smooth
function $u:M \times \lbrack 0,T)\rightarrow \mathbb{R},$ where $T\in
(0,\infty ],$ is a solution to the heat equation on $M $ if
\begin{equation}
\partial _{t}u=\Delta u.  \label{heat1}
\end{equation}%
One of the most important properties satisfied by the heat equation is the {%
maximum principle}, which says that for any smooth solution to the heat
equation, whatever point-wise bounds hold at $t=0$ also hold for $t>0$ \cite%
{CaoChow}. This property exhibits the smoothing behavior of the heat
diffusion (\ref{heat1}) on $M $.

Closely related to the heat diffusion (\ref{heat1}) is the (the Fields medal winning) Perelman entropy--action functional, which is on a $3n-$manifold $M $ with a Riemannian
metric $g_{ij}$ and a (temperature-like) scalar function $f$ given by \cite%
{Perel1}
\begin{equation}
\mathcal{E}=\int_{M }(R+|\nabla f|^{2}){\mathrm{e}}^{-f}d\mu   \label{F}
\end{equation}%
where $R$ is the scalar Riemann curvature on $M$, while $d\mu $ is the volume $3n-$form on $M$, defined as
\begin{equation}
d\mu =\sqrt{\det (g_{ij})}\,dx^{1}\wedge dx^{2}\wedge ...\wedge dx^{3n}.
\label{dmu}
\end{equation}

During the {Ricci flow} (\ref{RF})--(\ref{RFlim}) on the crowd manifold (\ref{crwdMan}), that is, during the inertial metric crowd flow, $C_{t}:t\mapsto (M(t),g(t))$, the Perelman entropy functional (\ref{F}) evolves as
\begin{equation}
\partial _{t}\mathcal{E}=2\int |R_{ij}+\nabla _{i}\nabla _{j}f|^{2}{\mathrm{e%
}^{-f}}d\mu {.}  \label{dF}
\end{equation}

Now, the \emph{crowd breathers} are solitonic crowd behaviors, which could be given by localized
periodic solutions of some nonlinear soliton PDEs, including the exactly
solvable sine--Gordon equation and the focusing nonlinear Schr\"{o}dinger
equation. In particular, the time--dependent crowd inertial metric $g_{ij}(t)$, evolving by the Ricci flow $g(t)$ given by (\ref{RF})--(\ref{RFlim}) on the crowd $3n-$manifold $M $ is the \emph{Ricci crowd
breather}, if for some $t_{1}<t_{2}$ and $\alpha >0$ the metrics $\alpha
g_{ij}(t_{1})$ and $g_{ij}(t_{2})$ differ only by a diffeomorphism; the
cases $\alpha =1,\alpha <1,\alpha >1$ correspond to steady, shrinking and
expanding crowd breathers, respectively. Trivial crowd breathers, for which
the metrics $g_{ij}(t_{1})$ and $g_{ij}(t_{2})$ on $M $ differ only by
diffeomorphism and scaling for each pair of $t_{1}$ and $t_{2}$, are the \emph{crowd
Ricci solitons}. Thus, if we consider the Ricci flow (\ref{RF})--(\ref{RFlim}) as a biodynamical system on
the space of Riemannian metrics modulo diffeomorphism and scaling, then
crowd breathers and solitons correspond to periodic orbits and fixed points
respectively. At each time the Ricci soliton metric satisfies on $M $
an equation of the form \cite{Perel1}
\begin{equation*}
R_{ij}+cg_{ij}+\nabla _{i}b_{j}+\nabla _{j}b_{i}=0,
\end{equation*}%
where $c$ is a number and $b_{i}$ is a 1--form; in particular, when $b_{i}=%
\frac{1}{2}\nabla _{i}a$ for some function $a$ on $M ,$ we get a
gradient Ricci soliton.

Define $\lambda (g_{ij})=\inf \mathcal{E}(g_{ij},f),$ where infimum is taken
over all smooth $f,$ satisfying
\begin{equation}
\int_{M }{\mathrm{e}^{-f}}d\mu =1.  \label{eDm}
\end{equation}%
$\lambda (g_{ij})$ is the lowest eigenvalue of the operator $-4\Delta +R.$
Then the entropy evolution formula (\ref{dF}) implies that $\lambda
(g_{ij}(t))$ is non-decreasing in $t,$ and moreover, if $\lambda
(t_{1})=\lambda (t_{2}),$ then for $t\in \lbrack t_{1},t_{2}]$ we have $%
R_{ij}+\nabla _{i}\nabla _{j}f=0$ for $f$ which minimizes $\mathcal{E}$ on $%
M $ \cite{Perel1}. Therefore, a steady breather on $M $ is
necessarily a steady soliton.

If we define the conjugate {heat operator} on $M $ as
\begin{equation*}
\Box ^{\ast }=-\partial /\partial t-\Delta +R
\end{equation*}%
then we have the {conjugate heat equation}:
~$
\Box ^{\ast }u=0.
$~

The entropy functional (\ref{F}) is nondecreasing under the coupled {%
Ricci--diffusion flow} on $M $ \cite{IvRicciSiam}
\begin{equation}
\partial _{t}g_{ij} =-2R_{ij}, \qquad
\partial _{t}u =-\Delta u+\frac{R}{2}u-\frac{|\nabla u|^{2}}{u},
\label{conHeat}
\end{equation}%
where the second equation ensures
~$
\int_{M }u^{2}d\mu =1,
$~
to be preserved by the Ricci flow $g(t)$ on $M $. If we define $\ u=%
\mathrm{e}^{-\frac{f}{2}}$, then (\ref{conHeat}) is equivalent to $f-$%
evolution equation on $M $ (the nonlinear backward heat equation),
\begin{equation*}
\partial _{t}f=-\Delta f+|\nabla f|^{2}-R,
\end{equation*}%
which instead preserves (\ref{eDm}). The coupled Ricci--diffusion flow (\ref%
{conHeat}) is the most general biodynamic model of the crowd reaction--diffusion
processes on $M$. In a recent study \cite{DifJennings} this general model has
been implemented for modelling a generic perception--action cycle with
applications to robot navigation in the form of a dynamical grid.

Perelman's functional $\mathcal{E}$ is analogous to negative thermodynamic
entropy \cite{Perel1}. Recall that thermodynamic {partition function} for a
generic canonical ensemble at temperature $\beta ^{-1}$ is given by
\begin{equation}
Z=\int \mathrm{e}^{{-\beta E}}d\omega (E),  \label{Z}
\end{equation}%
where $\omega (E)$ is a `density measure', which does not depend on $\beta .$
From it, the {average energy} is given by
~$
\left\langle E\right\rangle =-\partial _{\beta }\ln Z,
$
the {entropy }is
~$
S=\beta \left\langle E\right\rangle +\ln Z,
$
and the {fluctuation }is
~$
\sigma =\left\langle (E-\left\langle E\right\rangle )^{2}\right\rangle
=\partial _{\beta ^{2}}\ln Z.
$

If we now fix a closed $3n-$manifold $M $ with a probability measure $m$
and a metric $g_{ij}(\tau )$ that depends on the temperature $\tau $, then
according to equation
\begin{equation*}
\partial _{\tau }g_{ij}=2(R_{ij}+\nabla _{i}\nabla _{j}f),
\end{equation*}%
the partition function (\ref{Z}) is given by
\begin{equation}
\ln Z=\int (-f+\frac{n}{2})\, dm.  \label{lnZ}
\end{equation}%
From (\ref{lnZ}) we get (see \cite{Perel1})
\begin{eqnarray*}
&&\left\langle E\right\rangle  =-\tau ^{2}\int_{M}(R+|\nabla f|^{2}-\frac{n}{%
2\tau })\, dm,\qquad S=-\int_{M}(\tau (R+|\nabla f|^{2})+f-n)\, dm, \\
&&\sigma  =2\tau ^{4}\int_{M}|R_{ij}+\nabla _{i}\nabla _{j}f-\frac{1}{2\tau }%
g_{ij}|^{2}\, dm,\qquad\text{where~~}\, dm=u\,dV,~~ u=(4\pi \tau )^{-\frac{n}{2}}\mathrm{e}^{-f}.
\end{eqnarray*}

From the above formulas, we see that the fluctuation $\sigma $ is
nonnegative; it vanishes only on a gradient shrinking soliton. $\left\langle
E\right\rangle $ is nonnegative as well, whenever the flow exists for all
sufficiently small $\tau >0$. Furthermore, if the heat function $u$: (a)
tends to a $\delta -$function as $\tau \rightarrow 0,$ or (b) is a limit of
a sequence of partial heat functions $u_{i},$ such that each $u_{i}$ tends
to a $\delta -$function as $\tau \rightarrow \tau _{i}>0,$ and $\tau
_{i}\rightarrow 0,$ then the entropy $S$ is also nonnegative. In case (a),
all the quantities $\left\langle E\right\rangle ,S,\sigma $ tend to zero as $%
\tau \rightarrow 0,$ while in case (b), which may be interesting if $%
g_{ij}(\tau )$ becomes singular at $\tau =0,$ the entropy $S$ may tend to a
positive limit.

\subsection{Chaotic inter-phase in crowd dynamics induced by its Riemannian geometry change}

Recall that $\cal CD$ transition map (\ref{cd}) is defined by the chaotic
crowd phase-transition amplitude
\begin{equation*}
\left\langle \overset{\partial _{t}S=0}{\mathrm{PHYS.~ACTION}}\right\vert
CHAOS\left\vert \overset{\partial _{t}S>0}{\mathrm{MENTAL~PREP.}}%
\right\rangle :=\int_M \mathcal{D}[x]\,\mathrm{e}^{iA[x]},
\end{equation*}
where we expect the inter-phase chaotic behavior (see \cite{IAY}). To show that this chaotic inter-phase is caused by the change in Riemannian geometry of the crowd $3n-$manifold $M$, we will first simplify the $\cal CD$ action functional (\ref{Fey1}) as
\begin{equation}
A[x]={\frac{1}{2}}\int_{t_{ini}}^{t_{fin}}[g_{ij}\,\dot{x}^{i}\dot{x}%
^{j}-V(x,\dot{x})]\,dt,  \label{locAct}
\end{equation}%
with the associated standard Hamiltonian, corresponding to the amalgamate version of (\ref{ham1}),
\begin{equation}
H(p,x)=\sum_{i=1}^{N}\frac{1}{2}p_{i}^{2}+V(x,\dot{x}),  \label{Ham}
\end{equation}%
where $p_i$ are the SE(2)--momenta, canonically conjugate to the individual agents' SE(2)--coordinates $x^i,~(i=1,...,3n)$. Biodynamics of systems with action (\ref{locAct}) and
Hamiltonian (\ref{Ham}) are given by the set of \emph{geodesic equations}
\cite{GaneshSprBig,GaneshADG}
\begin{equation}
\frac{d^{2}x^{i}}{ds^{2}}+\Gamma _{jk}^{i}\frac{dx^{j}}{ds}\frac{dx^{k}}{ds}%
=0,  \label{geod-mot}
\end{equation}%
where $\Gamma _{jk}^{i}$ are the Christoffel symbols of the affine
Levi--Civita connection of the Riemannian $\cal CD$ manifold $M$.
In this geometrical framework, the instability of the trajectories is the
instability of the geodesics, and it is completely determined by the
curvature properties of the $\cal CD$ manifold $M$ according to
the {Jacobi equation} of geodesic deviation \cite{GaneshSprBig,GaneshADG}
\begin{equation}
\frac{D^{2}J^{i}}{ds^{2}}+R_{~jkm}^{i}\frac{dx^{j}}{ds}J^{k}\frac{dx^{m}}{ds}%
=0,  \label{eqJ}
\end{equation}%
whose solution $J$, usually called {Jacobi variation field}, locally
measures the distance between nearby geodesics; $D/ds$ stands for the {%
covariant derivative} along a geodesic and $R_{~jkm}^{i}$ are the components
of the {Riemann curvature tensor} of the $\cal CD$ manifold $M$.

The relevant part of the Jacobi equation (\ref{eqJ}) is given by the {%
tangent dynamics equation} \cite{pre96,CCC97}
\begin{equation}
\ddot{J}^{\,i}+R_{~0k0}^{i}J^{k}=0,\qquad (i,k=1,\dots ,3n),
\label{eqdintang}
\end{equation}%
where the only non-vanishing components of the curvature tensor of the
$\cal CD$ manifold $M$ are
\begin{equation}
R_{~0k0}^{i}=\partial ^{2}V/\partial x^{i}\partial x^{k}. \label{rok}
\end{equation}

The tangent dynamics equation (\ref{eqdintang}) can be used to define {%
Lyapunov exponents} in dynamical systems given by the Riemannian action (\ref%
{locAct}) and Hamiltonian (\ref{Ham}), using the formula \cite{physrep}
\begin{equation}
\lambda _{1}=\lim_{t\rightarrow \infty }1/2t\log (M
_{i=1}^{N}[J_{i}^{2}(t)+J_{i}^{2}(t)]/M
_{i=1}^{N}[J_{i}^{2}(0)+J_{i}^{2}(0)]).  \label{Lyap1}
\end{equation}%
Lyapunov exponents measure the {strength of dynamical chaos in the crowd
 behavior dynamics}. The sum of positive Lyapunov exponents defines the \emph{Kolmogorov--Sinai entropy}.

\subsection{Crowd nonequilibrium phase transitions induced by manifold\\ topology change}

Now, to relate these results to topological phase transitions within the
$\cal CD$ manifold $M$ given by (\ref{crwdMan}), recall that any two high--dimensional
manifolds $M _{v}$ and $M _{v^{\prime }}$ have the same topology
if they can be continuously and differentiably deformed into one another,
that is if they are diffeomorphic. Thus by {topology change} the `loss of
diffeomorphicity' is meant \cite{Pet07}. In this respect, the so--called {%
topological theorem} \cite{FP04} says that non--analyticity is the `shadow'
of a more fundamental phenomenon occurring in the system's configuration
manifold (in our case the $\cal CD$ manifold): a topology change within the
family of equipotential hypersurfaces
\begin{equation*}
M _{v}=\{(x^{1},\dots ,x^{3n})\in \mathbb{R}^{3n}|\ V(x^{1},\dots
,x^{3n})=v\},
\end{equation*}%
where $V$ and $x^{i}$ are the microscopic interaction potential and
coordinates respectively. This topological approach to PTs stems from the
numerical study of the dynamical counterpart of phase transitions, and
precisely from the observation of discontinuous or cuspy patterns displayed
by the largest Lyapunov exponent $\lambda _{1}$ at the {transition energy}
\cite{physrep}. Lyapunov exponents cannot be measured in laboratory
experiments, at variance with thermodynamic observables, thus, being genuine
dynamical observables they are only be estimated in numerical simulations of
the microscopic dynamics. If there are critical points of $V$ in
configuration space, that is points $x_{c}=[{\overline{x}}_{1},\dots ,{%
\overline{x}}_{3n}]$ such that $\left. \nabla V(x)\right\vert _{x=x_{c}}=0$,
according to the {Morse Lemma} \cite{hirsch}, in the neighborhood of any {%
critical point} $x_{c}$ there always exists a coordinate system {$x$}$(t)=[${%
$x$}$^{1}(t),...,${$x$}$^{3n}(t)]$ for which \cite{physrep}
\begin{equation}
V({x})=V(x_{c})-{x}_{1}^{2}-\dots -{x}_{k}^{2}+{x}_{k+1}^{2}+\dots +{x}%
_{3n}^{2},  \label{morsechart}
\end{equation}%
where $k$ is the {index of the critical point}, i.e., the number of negative
eigenvalues of the Hessian of the potential energy $V$. In the neighborhood
of a critical point of the $\cal CD$--manifold $M $, equation (\ref{morsechart})
yields the simplified form of (\ref{rok}),
~$
\partial ^{2}V/\partial x^{i}\partial x^{j}=\pm \delta _{ij},
$
giving $j$ unstable directions that contribute to the exponential
growth of the norm of the tangent vector $J$.

This means that the strength of dynamical chaos within the $\cal CD$--manifold $M $, measured by the largest Lyapunov exponent $\lambda
_{1}$ given by (\ref{Lyap1}), is affected by the existence of critical
points $x_{c}$ of the potential energy $V(x)$. However, as $V(x)$ is bounded
below, it is a good {Morse function}, with no vanishing eigenvalues of its
Hessian matrix. According to {Morse theory} \cite{hirsch}, the existence of
critical points of $V$ is associated with topology changes of the
hypersurfaces $\{M _{v}\}_{v\in \mathbb{R}}$.
The topology change of the $\{M _{v}\}_{v\in
\mathbb{R}}$ at some $v_{c}$ is a {necessary} condition for a phase
transition to take place at the corresponding energy value \cite{FP04}. The topology
changes implied here are those described within the framework of Morse
theory through `attachment of handles' \cite{hirsch} to the $\cal CD$--manifold $%
M$.

In our path--integral language this means that suitable
topology changes of equipotential submanifolds of the $\cal CD$--manifold $M $ can entail thermodynamic--like phase transitions
\cite{Haken1,Haken2,Haken3}, according to the general formula:
\begin{equation*}
\langle \mathrm{phase~out}\,|\,\mathrm{phase~in}\rangle\,:=
\int_{\rm top-ch}\mathcal{D}[w\Phi]\, {\rm e}^{iS[\Phi]}.
\end{equation*}
The statistical behavior of the crowd biodynamics system with the action functional (\ref{locAct}) and
the Hamiltonian (\ref{Ham}) is encompassed, in the canonical
ensemble, by its {partition function}, given by the Hamiltonian path
integral \cite{GaneshADG}%
\begin{equation}
Z_{3n}=\int_{\rm top-ch}\mathcal{D}[p]\mathcal{D}[x]\exp \{ \mathrm{i}\int_{t}^{t{%
^{\prime }}}[p_i\,\dot{x}^i-H(p,x)]\,d\tau \} ,  \label{PI1}
\end{equation}%
where we have used the shorthand notation
\begin{eqnarray*}
\int_{\rm top-ch}\mathcal{D}[p]\mathcal{D}[x]\equiv \int \prod_{\tau }\frac{dx(\tau )dp(\tau )%
}{2\pi }.
\end{eqnarray*}%
The path integral (\ref{PI1}) can be calculated as the
partition function \cite{FPS00},
\begin{eqnarray}
&&Z_{3n}(\beta )=\int \prod_{i=1}^{3n}dp_{i}\,dx^{i}\mathrm{e}^{-\beta
H(p,x)}=\left( \frac{\pi }{\beta }\right) ^{\frac{3n}{2}}\!\!\int
\prod_{i=1}^{3n}dx^{i}\mathrm{e}^{-\beta V(x)}  \notag \\
&&=\left( \frac{\pi }{\beta }\right) ^{\frac{3n}{2}}\int_{0}^{\infty }dv\,%
\mathrm{e}^{-\beta v}\int_{M _{v}}\frac{d\sigma }{\Vert \nabla V\Vert } , \label{zeta}
\end{eqnarray}%
where the last term is written using the so--called {co--area formula} \cite%
{federer}, and $v$ labels the equipotential hypersurfaces $M _{v}$ of
the $\cal CD$ manifold $M $,
\begin{equation*}
M _{v}=\{(x^{1},\dots ,x^{3n})\in \mathbb{R}^{3n}|V(x^{1},\dots
,x^{3n})=v\}.
\end{equation*}%
Equation (\ref{zeta}) shows that the relevant statistical information is
contained in the canonical configurational partition function
\begin{equation*}
Z_{3n}^{C}=\int \prod dx^{i}\,V(x)\,\mathrm{e}^{-\beta V(x)}.
\end{equation*}%
Note that $Z_{3n}^{C}$ is decomposed, in the last term of (\ref{zeta}), into
an infinite summation of geometric integrals,
\begin{equation*}
\int_{M _{v}}d\sigma /\Vert \nabla V\Vert ,
\end{equation*}%
defined on the $\{M _{v}\}_{v\in \mathbb{R}}$. Once the microscopic
interaction potential $V(x)$ is given, the configuration space of the system
is automatically foliated into the family $\{M _{v}\}_{v\in \mathbb{R}}$
of these equipotential hypersurfaces. Now, from standard statistical
mechanical arguments we know that, at any given value of the inverse
temperature $\beta $, the larger the number $3n$, the closer to $M
_{v}\equiv M _{u_{\beta }}$ are the microstates that significantly
contribute to the averages, computed through $Z_{3n}(\beta )$, of
thermodynamic observables. The hypersurface $M _{u_{\beta }}$ is the
one associated with
\begin{equation*}
u_{\beta }=(Z_{3n}^{C})^{-1}\int \prod dx^{i}V(x)\,\mathrm{e}^{-\beta V(x)},
\end{equation*}%
the average potential energy computed at a given $\beta $. Thus, at any $%
\beta $, if $3n$ is very large the effective support of the canonical measure
shrinks very close to a single $M _{v}=M _{u_{\beta }}$. Hence,
the basic origin of a phase transition lies in a suitable topology change of
the $\{M _{v}\}$, occurring at some $v_{c}$ \cite{FPS00}. This topology
change induces the singular behavior of the thermodynamic observables at a
phase transition. It is conjectured that the counterpart of a phase
transition is a breaking of diffeomorphicity among the surfaces $M _{v}$%
, it is appropriate to choose a {diffeomorphism invariant} to probe if and
how the topology of the $M _{v}$ changes as a function of $v$.
Fortunately, such a topological invariant exists, the {Euler characteristic}
of the crowd manifold $M $, defined by \cite{GaneshSprBig,GaneshADG}
\begin{equation}
\chi (M )=\sum_{k=0}^{3n}(-1)^{k}b_{k}(M ),  \label{chi}
\end{equation}%
where the {Betti numbers} $b_{k}(M )$ are diffeomorphism invariants ($b_{k}$ are the dimensions of the de Rham's cohomology
groups $H^{k}(M ;\mathbb{R})$; therefore the $b_{k}$ are
integers). This homological formula can be simplified by the use of the {%
Gauss--Bonnet theorem}, that relates $\chi (M )$ with the total {%
Gauss--Kronecker curvature} $K_{G}$ of the $\cal CD$--manifold $M $ given by \cite{GaneshADG,Complexity}
$$
\chi (M ) =\int_{M }K_{G}\, d\mu ,\qquad \text{where  $d\mu$  is given by (\ref{dmu}). }$$

\section{Conclusion}

Our understanding of crowd dynamics is presently limited in important ways; in particular, the lack of a geometrically \emph{predictive} theory of crowd behavior restricts the ability for authorities to intervene appropriately, or even to recognize when such intervention is needed. This is not merely an idle theoretical investigation: given increasing population sizes and thus increasing opportunity for the formation of large congregations of people, death and injury due to trampling and crushing -- even within crowds that have not formed under common malicious intent -- is a growing concern among police, military and emergency services. This paper represents a contribution towards the understanding of crowd behavior for the purpose of better informing decision--makers about the dangers and likely consequences of different intervention strategies in particular circumstances.

In this paper, we have proposed an entropic geometrical model of crowd dynamics, with dissipative kinematics, that operates across macro--, micro-- and meso--levels. This proposition is motivated by the need to explain the dynamics of crowds across these levels simultaneously. We contend that only by doing this can we expect to adequately characterize the geometrical properties of crowds with respect to regimes of behavior and the changes of state that mark the boundaries between such regimes.

In pursuing this idea, we have set aside traditional assumptions with respect to the separation of mind and body. Furthermore, we have attempted to transcend the long--running debate between contagion and convergence theories of crowd behavior with our multi-layered approach: rather than representing a reduction of the whole into parts or the emergence of the whole from the parts, our approach is build on the supposition that the direction of logical implication can and does flow in both directions simultaneously. We refer to this third alternative, which effectively unifies the other two, as \emph{behavioral composition.}

The most natural statistical descriptor is crowd entropy, which satisfies the extended second thermodynamics law applicable to open systems comprised of many components. Similarities between the configuration manifolds of individual (micro--level) and crowds (macro--level) motivate our claim that goal--directed movement operates under entropy conservation, while natural crowd dynamics operates under monotonically increasing entropy functions. Of particular interest is what happens between these distinct topological phases: the phase transition is marked by chaotic movement.

We contend that this approach provides a basis on which one can build a geometrically predictive model of crowd behavior dynamics -- over and above the existing approaches, which are largely explanatory.
The current paper develops an entropy formulation of crowd dynamics as a three-step process involving individual and collective behavior
dynamics, and –- crucially –- non-equilibrium phase transitions whereby the forces operating at the microscopic level result in geometrical change at the macroscopic level. We have incorporated both geometrical and algorithmic notions of entropy as well as chaos in studying the topological phase transition between the entropy conservation of physical action and the entropy increase during
internal, action preparation stages of behavior. Given these formulations, future research can focus on: (i) crowd simulations in 3D graphics environments, (ii) motivated cognition underpinning crowd dynamics and (iii) mechanisms of abrupt change in crowd behavior, such as crowd turbulence and flashpoints.\\

\noindent\textbf{Acknowledgment}\\

\noindent We thank Dr. A.C. Kalloniatis, DSTO Canberra, Fairbairn, A.C.T. for useful discussions about the use of complex-- and real--valued path integrals in relation to chaotic phase transitions in crowd modelling.

\section{Appendix}

General nonlinear stochastic dynamics, developed in a framework of Feynman
path integrals, have recently \cite{IA} been applied to Lewinian
field--theoretic psychodynamics \cite{Lewin97},
resulting in the development of a new concept of Life--Space
Foam (LSF) as a natural medium for motivational and cognitive
psychodynamics. According to the LSF--formalism, the classic
Lewinian life space can be macroscopically represented as a smooth
manifold with steady force--fields and behavioral paths, while at
the microscopic level it is more realistically represented as a
collection of wildly fluctuating force--fields, (loco)motion paths
and local geometries (and topologies with holes).

A set of least--action principles is used to model the smoothness
of global, macro--level LSF paths, fields and geometry, according
to the following prescription. The action $S[\Phi]$,
with dimensions of $Energy\times Time=Effort$ and
depending on macroscopic paths, fields and geometries (commonly
denoted by an abstract field symbol $\Phi^i$) is defined as a
temporal integral from the {initial} time instant $t_{ini}$ to the
{final} time instant $t_{fin}$,
\begin{equation}
S[\Phi]=\int_{t_{ini}}^{t_{fin}}\mathfrak{L}[\Phi]\,dt,
\label{act}
\end{equation}%
with {Lagrangian density} given by
\begin{equation*}
\mathfrak{L}[\Phi]=\int
d^{n}x\,\mathcal{L}(\Phi_i,\partial_{x^j}\Phi^i),
\end{equation*}%
where the integral is taken over all $n$ coordinates $x^j=x^j(t)$
of the LSF, and $\partial_{x^j}\Phi^i$ are time and space partial
derivatives of the $\Phi^i-$variables over coordinates. The
standard {least action principle}
\begin{equation}
\delta S[\Phi]=0, \label{actPr}
\end{equation}
gives, in the form of the so--called Euler--Lagrangian equations,
a shortest (loco)motion path, an extreme force--field, and a
life--space geometry of minimal curvature (and without holes). In
this way, we have
obtained macro--objects in the global LSF: a single path described
by Newtonian--like equation of motion, a single force--field
described by Maxwellian--like field equations, and a single
obstacle--free Riemannian geometry (with global topology without
holes).

To model the corresponding local, micro--level LSF structures of
rapidly fluctuating MD \& CD, an adaptive path integral is
formulated, defining a multi--phase and multi--path (multi--field
and multi--geometry) {transition amplitude} from the motivational state of
$Intention$ to the cognitive state of $Action$,
\begin{equation}
\langle Action|Intention\rangle_{total} :=\int\mathcal{D}[w\Phi]\,
{\mathrm e}^{\mathrm i S[\Phi]}, \label{pathInt1}
\end{equation}
where the Lebesgue integration is performed over all continuous
$\Phi^i_{con}=paths+fields+geometries$, while summation is
performed over all discrete processes and regional topologies
$\Phi^j_{dis}$. The symbolic differential $\mathcal{D}[w\Phi]$ in
the general path integral (\ref{pathInt}) represents an {adaptive
path measure}, defined as a weighted product
\begin{equation}
\mathcal{D}[w\Phi]=\lim_{N\to\infty}\prod_{s=1}^{N}w_sd\Phi
_{s}^i, \qquad ({i=1,...,n=con+dis}).\label{prod1}
\end{equation}
The adaptive path integral (\ref{pathInt1})--(\ref{prod1})
represents an $\infty-$dimensional neural network, with weights
$w$ updating by the general rule \cite{NeuFuz}
$$new\;value(t+1)\; =\; old\;value(t)\;+\; innovation(t).$$

\end{document}